\documentclass[12pt]{elsart-hep}
\usepackage{epsfig,graphics}

\newcommand{\tr}{\rm tr \,}

\newcommand{\one}{\bf 1}

\newcommand{\barqslash}{\FMslash {\bar q}}

\newcommand{\qslash}{\FMslash q}

\newcommand{\Partialslash}{\FMslash \partial}

\newcommand{\TableHeader}[0]{\begin{tabular}{|c|c|c|c|c|c|} \hline (I,S)  &  Kanal &  Lage &  |g|  & Lage  &|g| \\}
\def\rescale{\fontsize{6}{1}}

\begin{document}
\begin{frontmatter}
\title{Coupled-channel study of \\crypto-exotic baryons with charm}
\author[GSI]{J. Hofmann}
\author[GSI]{and M.F.M. Lutz}
\address[GSI]{Gesellschaft f\"ur Schwerionenforschung (GSI)\\
Planck Str. 1, 64291 Darmstadt, Germany}
\begin{abstract}
Identifying a zero-range exchange of vector
mesons as the driving force for the s-wave scattering of pseudo-scalar mesons
off the  baryon ground states, a rich spectrum of molecules
is formed. We argue that chiral symmetry and
large-$N_c$ considerations determine that part of the interaction which generates the spectrum.
A bound state with exotic quantum numbers is predicted at mass 2.78 GeV. It
couples strongly to the $(\bar D_s\,N),(\bar D \,\Lambda ),(\bar D\,\Sigma)$ channels.
A further charm minus-one system is predicted at mass 2.84 GeV as a result of
$(\bar D_s \Lambda), (\bar D\,\Xi)$ interactions.
We suggest the existence of strongly bound crypto-exotic baryons, which
contain a charm-anti-charm pair. Such states are narrow since they
can decay only via OZI-violating processes. A narrow nucleon resonance is found at mass
3.52 GeV. It is a coupled-channel bound state of the $(\eta_c\,N), (\bar D\,\Sigma_c)$
system, which decays dominantly into the $(\eta' N)$ channel. Furthermore two isospin singlet
hyperon states at mass 3.23 GeV and 3.58 GeV are observed as a consequence
of coupled-channel interactions of the $(\bar D_s\,\Lambda_c),
(\bar D\,\Xi_c)$ and $(\eta_c \,\Lambda),(\bar D\,\Xi_c')$ states. Most striking is the
small width of about 1 MeV of the lower state. The upper state may be significantly broader
due to a strong coupling to the $(\eta' \Lambda)$ state. The spectrum of
crypto-exotic charm-zero states is completed with an isospin triplet state at 3.93 GeV and an
isospin doublet state at 3.80 GeV. The dominant decay modes involve again the $\eta'$ meson.
The two  so far observed s-wave baryons with charm one are
recovered. We argue that the $\Lambda_c(2880)$ is not a s-wave state.
In addition to those states we predict the existence of about ten narrow s-wave
baryon states with masses below 3 GeV. A triplet of crypto-exotic states decaying
dominantly into channels with an $\eta'$ is obtained with masses 4.24 GeV and 4.44 GeV.
In the charm-two sector we predict in addition to the
chiral excitations of the ground states two triplets of bound states formed by channels involving
open-charm mesons. The binding energy of the latter is larger than the one of the chiral
excitations.
\end{abstract}
\end{frontmatter}

\section{Introduction}

The existence of strongly bound crypto-exotic baryon systems with
hidden charm would be a striking feature of strong interactions
\cite{Brodsky:Schmidt:Teramond:90,Kaidalov:Volkovitsky:92,Gobbi:Riska:Scoccola:92}. Such states
may be narrow since their strong decays are OZI-suppressed \cite{Landsberg:94}.
There are experimental hints that such states may indeed be part of nature. A high statistics bubble
chamber experiment performed 30 years ago with a $K^-$ beam reported on a possible signal for
a hyperon resonance of mass 3.17 GeV of width smaller than 20 MeV \cite{Amirzadeh:79}.
About ten
years later a further bubble chamber experiment using a high energy $\pi^-$ beam suggested
a nucleon resonance of mass 3.52 GeV with a narrow width of $7^{+20}_{-7}$ MeV \cite{Karnaukhov:91}.

It is the purpose of the present work to perform a study addressing the
possible existence of crypto-exotic baryon systems. In view of the highly speculative nature
of such states it is important to correlate the properties of such states to those firmly
established, applying a unified and quantitative framework. Our strategy will be to extend
previous works \cite{LK04-charm,LK05} that performed a coupled-channel study of
the s-wave scattering processes where a Goldstone boson hits an open-charm baryon ground state.
The spectrum of $J^P=\frac{1}{2}^-$ and $J^P=\frac{3}{2}^-$ molecules obtained in
\cite{LK04-charm,LK05} is quite compatible with the so
far very few observed states. We mention that analogous computations
successfully describe the spectrum of open-charm mesons with $J^P=0^+$ and $1^+$
quantum numbers \cite{KL04,HL04}. These developments were driven by the conjecture
that meson and baryon resonances that do not belong to the
large-$N_c$ ground state of QCD should be viewed as hadronic molecular states
\cite{LK02,LWF02,LK04-axial,Granada,Copenhagen}. Extending those computations to include
D- and $\eta_c$-mesons in the intermediate states offers the possibility to address the
formation of crypto-exotic baryon states
(see also \cite{Mo:80,Rho:Riska:Scoccola:92,Min:Oh:Rho:95,Oh:Kim:04}).
The results of \cite{LK04-charm,LK05} were based on the
leading order chiral Lagrangian, that predicts unambiguously the s-wave interaction
strength of Goldstone bosons with open-charm baryon states in terms of the pion decay constant.
Including the light vector mesons as explicit degrees of freedom in a chiral
Lagrangian gives an interpretation of the leading order interaction in terms of the zero-range
t-channel exchange of light vector mesons \cite{Weinberg:68,Wyld,Dalitz,sw88,Bando:85}. The latter
couple universally to any matter field in this type of approach. Given the assumption that the interaction
strength of D- and $\eta_c$-mesons with the baryon ground states is also dominated by the
t-channel exchange of the light vector mesons,  we are in a position to perform a quantitative
coupled-channel study of crypto-exotic baryon resonances.

The work is organized as follows. In section 2 the t-channel force implied by the exchange
of vector mesons is constructed defining the leading interaction of the mesons with the
baryons. We consider the ground states with $J^P= 0^-,\frac{1}{2}^+$ quantum numbers composed
out of $u,d,s,c$ quarks. Constraints imposed by chiral symmetry and large-$N_c$ QCD are
discussed. In section 3 we identify the coupled-channel states considered in our study.
Applying the formalism developed in \cite{LK02,LK04-axial} the scattering amplitudes are
written down explicitly in terms of Clebsch-Gordon coefficients collected in Appendix A.
Numerical results are presented in section 4. A brief summary is given in section 5.

The interaction strengths of the channels
that drive the resonance generation are predicted by chiral and large-$N_c$ properties of QCD.
The spectrum of $J^P=\frac{1}{2}^-$ molecules obtained is amazingly rich of structure.
We do recover a nucleon resonance at around 3.3-3.5 GeV as suggested in \cite{Karnaukhov:91}.
It is a coupled-channel bound-state of the $\eta_c\,N $ and $ \bar D\,\Sigma_c$
states, that decays dominantly into the $\eta' N$ channel. Its decay width is driven by the t-channel
exchange of charm. Using a SU(4) estimate a width results of about 150 MeV.
Moderate SU(4) breaking allows to reduce the width down to 5-20 MeV within the range suggested
in \cite{Karnaukhov:91}. A small OZI violating $\phi_\mu \,D \bar D $ vertex can be used to increase the
mass of the state up to the empirical value of 3.52 GeV. Reducing
the strength of the $\eta'$ coupling strength to the open-charm mesons, away from its SU(4) estimate,
the crypto-exotic nucleon resonance can be made with narrow width. The state belongs to a SU(3)
octet, with all members decaying preferably into channels involving the $\eta'$ meson. The isospin
singlet, doublet and triplet come at (3.58, 3.80, 3.93) GeV respectively.
Most striking is
an even stronger bound $(\bar D_s\Lambda_c),(\bar D\,\Xi_c)$ system. Its mass is close to
a state possibly observed at 3.17 GeV as reported in \cite{Amirzadeh:79}. Since it is a SU(3) singlet
its decay into channels involving the $\eta'$ meson is suppressed. As a result a width smaller
than 1 MeV is predicted. This is compatible with \cite{Amirzadeh:79}.

Further interesting results are obtained in the charm minus one sector. Exotic states
with strangeness minus one and two are predicted at mass 2.78 GeV and 2.84 GeV.
In the charm one sector we recover the $\Lambda_c(2593)$ as a narrow state coupling
strongly to the $(D\,N)$ and $(D_s\,\Lambda)$ states. The $\Xi_c(2790)$ is interpreted as a
bound state of the $(\bar K \,\Sigma_c),(\eta \,\Xi'_c)$ system. We argue that the
$\Lambda_c(2880)$ discovered by the CLEO collaboration can not be a s-wave state.
About ten additional narrow s-wave states are predicted in this sector with masses below 3 GeV.
For instance, narrow $\Xi_c(2670)$ and $\Xi_c(2760)$ stats are obtained that couple
strongly to the $(D\,\Sigma),(D_s\Xi)$ and $(D\,\Lambda),(D_s\Xi)$ states respectively.
A narrow $\Lambda_c(2815)$ state  is foreseen that
is formed by the $(\eta \,\Lambda_c), (K\,\Xi_c)$ system. A narrow $\Sigma_c(2620)$ state
couples strongly to the $(D\,N)$ and $(D_s\Sigma)$ states.

\newpage

\section{Coupled-channel interactions}

We study the interaction of pseudoscalar mesons with the ground-state baryons composed
out of u,d,s,c quarks. The pseudoscalar mesons that are considered in this work can be
grouped into multiplet fields $\Phi_{[9]}, \Phi_{[\bar 3]}$ and $\Phi_{[1]}$. We
introduce the fields
\begin{eqnarray}
&& \Phi_{[9]}=
\tau \cdot \pi(139) + \alpha^{\dag} \cdot K(494)
+ K^{\dag}(494) \cdot \alpha + \eta(547)\ \lambda_8
+ \sqrt{{\textstyle{2\over 3}}}\,{\one } \,\eta'(958)\,,
\nonumber\\
&& \Phi_{[\bar 3]} = {\textstyle{1\over \sqrt{2}}}\,\alpha^\dagger \cdot D_{\phantom{\mu}}(1867)
- {\textstyle{1\over \sqrt{2}}}\,D^t_{\phantom{\mu}}(1867)\cdot \alpha   +  i\,\tau_2\,D_{\phantom{\mu}}^{(s)}(1969)
\,,
\nonumber\\
&&\Phi_{[1]} = \eta_c(2980)\,,
\label{mesons}
\end{eqnarray}
decomposed further into SU(2) multiplets. The approximate masses in units
of MeV as used in this work are recalled in brackets \cite{PDG04}. The
matrices $\tau $ and $\alpha$ are given in terms of
the Gell-Mann SU(3) generators $\lambda_{1,...,8}$ with
\begin{eqnarray}
\tau = (\lambda_1,\lambda_2,\lambda_3), \quad \alpha^{\dagger}=
\frac1{\sqrt2}(\lambda_4+i\lambda_5,\lambda_6+i\lambda_7)\,.
\end{eqnarray}
The baryon states are collected into SU(3) multiplet fields $B_{[8]}, B_{[6]}, B_{[3]}$ and
$B_{[\bar 3]}$. Again a further decomposition into isospin multiplets is useful
\begin{eqnarray}
&& \sqrt{2}\,B_{[8]} =
 \tau \cdot \Sigma(1193) + \alpha^{\dag} \cdot N(939) +
  \Xi^{t}(1318) \cdot \alpha + \Lambda(1116)\ \lambda_8 \,,
\nonumber\\
&& \sqrt{2}\, B_{[6]} =
{\textstyle{1\over \sqrt{2}}}\,\alpha^{\dag}\cdot \Xi_c(2576)
+ {\textstyle{1\over \sqrt{2}}}\, \Xi_c^t (2576)\cdot \alpha +\Sigma_c(2452)\cdot (i\,\tau \,\tau_2)
\nonumber\\
&& \qquad \;\;\;\;\; + {\textstyle{\sqrt{2}\over 3}}\,
(1-\sqrt{3}\,\lambda_8)\,\Omega_c(2698) \,,
\nonumber\\
&& \sqrt{2}\, B_{[\bar 3]} =
{\textstyle{1\over \sqrt{2}}}\,\alpha^\dagger \cdot \Xi_{c\,}(2469)
- {\textstyle{1\over \sqrt{2}}}\,\Xi^t_{c\,}(2469)\cdot \alpha
+  i\,\tau_2\,\Lambda_{c}^{}(2285) \,,
\nonumber\\
&& \sqrt{2}\, B_{[3]} =
{\textstyle{1\over \sqrt{2}}}\,\alpha^\dagger \cdot \Xi_{cc}(3440)
- {\textstyle{1\over \sqrt{2}}}\,\Xi^t_{cc}(3440)\cdot \alpha   +
 i\,\tau_2\,\Omega_{cc}(3560) \,.
\label{baryons}
\end{eqnarray}
The mass of the $\Xi_{cc}$ state is unknown so far. We assume the somewhat ad-hoc value
3.44 GeV \footnote{There are hints that the double-charm $\Xi_{cc}$ state at 3519 MeV \cite{SELEX}
has not $J^P=\frac{1}{2}^+ $ quantum numbers \cite{SELEX:Russ,SELEX:Engelfried-a}.}.
The mass of the $\Omega_{cc}$ state is estimated from quark-model
calculations \cite{Koerner:94} which predict a typical splitting to the $\Xi_{cc}$ of
about 100-150 MeV. In the SU(3) limit the kinetic terms of the various fields take the form
\begin{eqnarray}
{\mathcal L}^{\rm{SU(3)}}_{\rm{kin}} &=&{\textstyle{1\over 4}}\,\tr \Big(
(\partial_\mu \,\Phi_{[9]})\,(\partial^\mu \,\Phi_{[9]})
 - m_{[9]}^2\,\Phi_{[9]}\,\Phi_{[9]} \Big)
 + \tr \bar B_{[8]}\,(i\,\Partialslash - M_{[8]})\,B_{[8]}
 \nonumber\\
 &+& \tr \bar B_{[6]}\,(i\,\Partialslash - M_{[6]})\,B_{[6]}
 +{\textstyle{1\over 2}}\,\tr \Big(
(\partial_\mu \,\Phi^\dagger_{[\bar 3]})\,(\partial^\mu \,\Phi_{[\bar 3]})
 - m_{[\bar 3]}^2\,\Phi^\dagger_{[\bar 3]}\,\Phi_{[\bar 3]} \Big)
  \nonumber\\
 &+&  \tr \bar B_{[\bar 3]}\,(i\,\Partialslash - M_{[\bar 3]})\,B_{[\bar 3]}
+\tr \bar B_{[3]}\,(i\,\Partialslash - M_{[3]})\,B_{[3]}
 \nonumber\\
 &+&  {\textstyle{1\over 2}}\,\Big(
(\partial_\mu \,\Phi_{[1]})\,(\partial^\mu \,\Phi_{[1]})
 - m_{[1]}^2\, \Phi_{[1]}\,\Phi_{[1]} \Big) \,,
\end{eqnarray}
where the mass parameters may be taken as appropriate averages of the values
implicit in (\ref{mesons}, \ref{baryons}).

In a first step we construct the interaction of the mesons and baryon fields introduced in
(\ref{mesons}, \ref{baryons}) with the nonet-field of light vector mesons
\begin{eqnarray}
&& V^{[9]}_\mu = \tau \cdot \rho_\mu(770)
+ \alpha^\dagger \cdot  K_\mu(894) +  K_\mu^\dagger(894) \cdot \alpha
\nonumber\\
&& \qquad +
\Big( {\textstyle{2\over 3}}+ {\textstyle{1\over \sqrt{3}}} \,\lambda_8  \Big)\, \omega_\mu(783)
+\Big(  {\textstyle{\sqrt{2}\over 3}}-\sqrt{{\textstyle{2\over 3}}}\,\lambda_8\Big)\,
\phi_\mu(1020) \,.
\label{light-vectors}
\end{eqnarray}
We write down a complete list of SU(3) invariant 3-point vertices that involve the minimal number
of derivatives. Consider first the terms involving pseudo-scalar fields:
\begin{eqnarray} \label{LM}
\mathcal{L}^{\rm{SU(3)}}_{\rm{int}}
&=& {\textstyle{i\over 4}}\, h_{\bar 3 \bar 3}^9\    {\rm tr}\left(
 (\partial_{\mu}\Phi_{[\bar 3]})\, \Phi^\dagger_{[\bar 3]} V_{[9]}^{\mu} -
  \Phi_{[\bar 3]}\,
(\partial_{\mu}\Phi^\dagger_{[\bar 3]})\,
V_{[9]}^{\mu}\right)
\nonumber\\
&+& {\textstyle{i\over 4}}\, h_{\bar 3 \bar 3}^1\    {\rm tr}\left(
(\partial_{\mu}\Phi_{[\bar 3]})\, \Phi^\dagger_{[\bar 3]}-
\Phi_{[\bar 3]}\,(\partial_{\mu}\Phi^\dagger_{[\bar 3]})  \right) \cdot {\rm tr} \left (V_{[9]}^{\mu} \right)
\nonumber\\
&+& {\textstyle{i\over 4}}\, h_{99}^9\    {\rm tr}\left(
(\partial_{\mu}\Phi_{[9]})\, \Phi_{[9]}\, V_{[9]}^{\mu}
-\Phi_{[9]}\,(\partial_{\mu} \Phi_{[9]})\,  V_{[9]}^{\mu} \right)\,.
\label{vector-pseudoscalar}
\end{eqnarray}
It needs to be emphasized that the terms in (\ref{vector-pseudoscalar}) are
not at odds with the constraints set by chiral symmetry provided the light vector
mesons are coupled to matter fields via a gauge principle \cite{Weinberg:68,Bando:85}.
The latter requires a correlation of the coupling constants $h$ in (\ref{vector-pseudoscalar})
\begin{eqnarray}
h_{99}^9 = \frac{(m^{(V)}_{[9]})^2}{2\,g\,f^2}\,, \qquad h_{\bar 3 \bar 3}^{9} = 2\,g \,,
\label{chiral-constraint-mesons}
\end{eqnarray}
with the pion decay constant $f \simeq 92$ MeV. Here the universal vector coupling strength is
$g \simeq 6.6$ and the mass of the light vector mesons is $m_{[9]}^{(V)}$.

We continue with the construction of the three-point vertices involving baryon fields.
A complete list of SU(3) invariant terms reads:
\begin{eqnarray}
\mathcal{L}^{\rm{SU(3)}}_{\rm{int}}
&=& {\textstyle{1\over 2}}\,g_{33}^{9}\,
{\tr }\Big(\bar B_{[3]}\,\gamma_\mu\, B_{[3]}  \,V_{[9]}^\mu\Big)
+{\textstyle{1\over 2}}\,g_{33}^{1}\,
{\tr }\Big(\bar B_{[3]}\,\gamma_\mu \, B_{[3]} \Big)\, {\tr} \Big( \,V_{[9]}^\mu \Big)
\nonumber\\
&+&{\textstyle{1\over 2}}\,g_{\bar 3 \bar 3}^{9}\,
{\tr }\Big(\bar B_{[\bar 3]}\,\gamma_\mu \,V_{[9]}^\mu\, B_{[\bar 3]} \Big)
+{\textstyle{1\over 2}}\,g_{\bar 3 \bar 3}^{1}\,
{\tr }\Big(\bar B_{[\bar 3]}\,\gamma_\mu \, B_{[\bar 3]} \Big)\, {\tr} \Big( \,V_{[9]}^\mu \Big)
\nonumber\\
&+&{\textstyle{1\over 2}}\,g_{66}^{9}\,
{\tr }\Big(\bar B_{[6]}\,\gamma_\mu \,V_{[9]}^\mu\, B_{[6]} \Big)
+{\textstyle{1\over 2}}\,g_{66}^{1}\,
{\tr }\Big(\bar B_{[6]}\,\gamma_\mu \, B_{[6]} \Big)\, {\tr} \Big( \,V_{[9]}^\mu \Big)
\nonumber\\
&+&{\textstyle{1\over 2}}\,g_{88}^{9-}\,
{\tr }\Big(\bar B_{[8]}\,\gamma_\mu \,\Big[ V_{[9]}^\mu\,, B_{[8]}\Big]_- \Big)
+{\textstyle{1\over 2}}\,g_{88}^{9+}\,
{\tr }\Big(\bar B_{[8]}\,\gamma_\mu \,\Big[ V_{[9]}^\mu\,, B_{[8]}\Big]_+ \Big)
\nonumber\\
&+&{\textstyle{1\over 2}}\,g_{88}^{1}\,
{\tr }\Big(\bar B_{[8]}\,\gamma_\mu \, B_{[8]} \Big)\, {\tr} \Big( \,V_{[9]}^\mu \Big)
\nonumber\\
&+&{\textstyle{1\over 2}}\,g_{\bar36}^{9}\,
{\tr }\Big(\bar B_{[\bar3]}\,\gamma_\mu\, V_{[9]}^\mu \, B_{[6]}+ \bar B_{[6]}\,\gamma_\mu\, V_{[9]}^\mu \, B_{[\bar 3]} \Big)\,.
\label{vector-baryons}
\end{eqnarray}
Within the hidden local symmetry model \cite{Bando:85} chiral
symmetry is recovered with
\begin{eqnarray}
&& g_{\bar 3 \bar 3}^9 =g_{66}^{9}=-g_{33}^{9}= 2\,g\,,
\qquad g_{88}^{9,-}= g\,, \qquad \! g_{88}^{9,+}= 0\,, \qquad g_{\bar36}^{9}=0 \,.
\label{chiral-constraint-baryons}
\end{eqnarray}
It is acknowledged that chiral symmetry does not constrain the coupling
constants in (\ref{vector-pseudoscalar}, \ref{vector-baryons}) involving the SU(3) singlet
part of the fields. The latter can, however, be
constrained  by a large-$N_c$ operator analysis \cite{DJM}.
At leading order in the $1/N_c$ expansion the OZI rule \cite{OZI} is predicted.  As
a consequence the estimates
\begin{eqnarray}
&& h^1_{\bar 3\bar 3} = -g\,,\qquad
 g^1_{\bar 3 \bar 3}=g^1_{66}=0 \,,\qquad  g^1_{88}=g^1_{3 3}= g\,,\quad
\label{OZI-constraint}
\end{eqnarray}
follow. We emphasize that the combination of chiral and large-$N_c$ constraints
(\ref{chiral-constraint-mesons}, \ref{chiral-constraint-baryons}, \ref{OZI-constraint})
determine all coupling constants introduced in (\ref{vector-pseudoscalar}, \ref{vector-baryons}).

Before constructing the t-channel exchange forces it is rewarding to dwell a bit more
on the coupling constants introduced in (\ref{mesons}, \ref{baryons}). One may wonder whether
the results obtained are compatible with the approximate SU(4) symmetry of QCD that would
arise in the limit of a light charm quark mass. As an amusing result of this exercise we will
demonstrate that the KSFR relation \cite{KSFRa,KSFRb,Djukanovic} can be derived by insisting
on SU(4) symmetric coupling constants. The SU(4) symmetric generalization of (\ref{mesons})
is readily written down
\begin{eqnarray}
&&\!{\mathcal L}^{\rm{SU(4)}}_{\rm{int}} = {\textstyle{i\over 4}}\,g\,{\tr}\Big(
 \big[  (\partial_\mu\,\Phi_{[16]})\,,\Phi_{[16]} \big]_- V^\mu_{[16]}\Big) \,,
\nonumber\\ \nonumber\\
&&\!\Phi_{[16]}\!=\!\left( \begin{array}{cccc}
\pi_0 \!+\! \frac{1}{\sqrt{3}}\,\eta + \!\sqrt{\frac{2}{3}}\,\eta'  &  \sqrt2\ \pi_+  &  \sqrt2\ K_+  &  \sqrt2\ \bar D^0  \\
\sqrt2\ \pi_-  &  -\pi_0 \!+ \!\frac{1}{\sqrt{3}}\,\eta +\! \sqrt{\frac{2}{3}}\,\eta'  &  \sqrt2\ K^0  &  -\sqrt2\ \bar D^-  \\
\sqrt2\ \bar K^-  &  \sqrt2\ \bar K^0  &  -\frac{2}{\sqrt{3}}\,\eta + \!\sqrt{\frac{2}{3}}\,\eta'  &  \sqrt2\ \bar D^-_s  \\
\sqrt2\ D^0  & - \sqrt2\ D^+  &  \sqrt2\ D^+_s  &  \sqrt{2}\,\eta_c \end{array} \right),
\nonumber\\
&&\!V^\mu_{[16]}\!=\!\left( \begin{array}{cccc}
\rho_0^\mu + \omega^\mu  &  \sqrt2\ \rho_+^\mu  &  \sqrt2\ K_{+}^\mu  &  \sqrt2\ \bar D_{0}^\mu  \\
\sqrt2\ \rho_-^\mu  &  -\rho_0^\mu + \omega^\mu   &  \sqrt2\ K_{0}^\mu  &  -\sqrt2\ \bar D_{-}^\mu  \\
\sqrt2\ \bar K_{-}^\mu  &  \sqrt2\ \bar K_{*0}^\mu  &  \sqrt2 \ \phi^\mu  &  \sqrt2\ \bar D_{-}^{s,\mu}  \\
\sqrt2\ D_{0}^\mu  &  -\sqrt2\ D_{+}^\mu &  \sqrt2\ D_{+}^{s,\mu}  &  \sqrt2 \ J/\Psi^\mu \end{array} \right).
\label{mesons-SU4}
\end{eqnarray}
We observe that the interaction (\ref{mesons-SU4}) is compatible with
(\ref{chiral-constraint-mesons}) and (\ref{OZI-constraint}) if and only if the
KSFR relation
\begin{eqnarray}
\frac{(m^{(V)}_{[9]})^2}{2\,f^2\,g} = g \,,
\label{KSFR-relation}
\end{eqnarray}
holds. This is a surprising result since at first sight the KSFR relation does not
seem to be connected with the physics of charm quarks. It is instructive to construct also the
SU(4) symmetric generalization of the interaction (\ref{vector-baryons}). The baryons
form a 20-plet in SU(4). Its field is represented by a tensor $B^{ijk}$, which is antisymmetric
in the first two indices. The indices $i,j,k$ run from one to four, where one can read off the
quark content of a baryon state by the identifications $1 \leftrightarrow u,
2 \leftrightarrow d, 3 \leftrightarrow s, 4 \leftrightarrow c$.
We write:
\begin{eqnarray}
&&{\mathcal L}^{\rm{SU(4)}}_{\rm{int}} = {\textstyle{1\over 4}}\,g \sum_{i,j,k,l=1}^4\,
\bar B^{[20]}_{ijk} \,\gamma^\mu\,\Big( V_{\mu,\,l}^{[16],k}\,B^{ijl}_{[20]}
+ 2\, V_{\mu,\,l}^{[16],j}\,B^{ilk}_{[20]}\Big)
\label{baryons-SU4} \\
&& \renewcommand{\arraystretch}{1.3}
\begin{array}{lll}B_{[20]}^{121}=p&         B_{[20]}^{122}=n&           B_{[20]}^{132}=\frac1{\sqrt2} \Sigma^0 -\frac1{\sqrt6}\Lambda\\*
    B_{[20]}^{213}=\frac2{\sqrt6}\Lambda &  B_{[20]}^{231}=\frac1{\sqrt2}\Sigma^0+\frac1{\sqrt6}\Lambda & B_{[20]}^{232}=\Sigma^-\\*
    B_{[20]}^{233}=\Xi^-&           B_{[20]}^{311}=\Sigma^+&        B_{[20]}^{313}=\Xi^0 \\
    \nonumber\\
    B_{[20]}^{141}=-\Sigma_c^{++}&  B_{[20]}^{142}=\frac1{\sqrt2}\Sigma_c^++\frac1{\sqrt6}\Lambda_c&  B_{[20]}^{143}=\frac1{\sqrt2}\Xi_c'^+-\frac1{\sqrt6}\Xi_c^+\\*
    B_{[20]}^{241}=\frac1{\sqrt2}\Sigma_c^+-\frac1{\sqrt6}\Lambda_c&  B_{[20]}^{242}=\Sigma_c^0&  B_{[20]}^{243}=\frac1{\sqrt2}\Xi_c'^0+\frac1{\sqrt6}\Xi_c^0\\*
    B_{[20]}^{341}=\frac1{\sqrt2}\Xi_c'^++\frac1{\sqrt6}\Xi_c^+&        B_{[20]}^{342}=\frac1{\sqrt2}\Xi_c'^0-\frac1{\sqrt6}\Xi_c^0&  B_{[20]}^{343}=\Omega_c\\
    \nonumber\\
    B_{[20]}^{124}=\frac2{\sqrt6}\Lambda_c^0&B_{[20]}^{234}=\frac2{\sqrt6}\Xi_c^0&  B_{[20]}^{314}=\frac2{\sqrt6}\Xi_c^+  \\
    \nonumber\\
    B_{[20]}^{144}=\Xi_{cc}^{++}&   B_{[20]}^{244}=-\Xi_{cc}^+&  B_{[20]}^{344}=\Omega_{cc} \,.
    \end{array}
\end{eqnarray}
It is pointed out that the relations (\ref{chiral-constraint-baryons}, \ref{OZI-constraint})
follow from the form of the interaction (\ref{baryons-SU4}).
For completeness we spell out the kinetic term written in terms of the
SU(4) multiplet fields:
\begin{eqnarray}
{\mathcal L}^{\rm{SU(4)}}_{\rm{kin}} &=& {\textstyle{1\over 4 }}\,\sum_{i,j=1}^4
\Big( (\partial_\mu\,\Phi^{i}_{[16],\,j})\,
(\partial^\mu\,\Phi^j_{[16],\,i}) - m^2_{[16]}\,\Phi^i_{[16],\,j}\,\Phi^j_{[16],\,i}\Big)
\nonumber\\
&+& {\textstyle{1\over 2 }}\,\sum_{i,j,k=1}^4 \,
\bar B^{[20]}_{ijk}\,\Big( i\,\Partialslash - M_{[20]}\Big) \,B_{[20]}^{ijk} \,.
\label{kinetic-SU4}
\end{eqnarray}
In contrast to the coupling constants the assumption of a SU(4) symmetry is
quite nonsensical for the meson and baryon masses \cite{Mathur:Okubo:Borchardt:75}.
The notion of $m_{[16]}$ and
$M_{[20]}$ in (\ref{kinetic-SU4}) is introduced only to illustrate the normalization of
the fields.

We close this section by investigating the coupling of heavy vector mesons, $V^\mu_{[\bar 3]}$ and
$V_{[1]}^\mu$ to the meson and baryon fields. The implications of the
vertices (\ref{mesons-SU4}) and (\ref{baryons-SU4}) is worked out. The multiplet fields are
introduced with
\begin{eqnarray}
&& V_\mu^{[\bar 3]} ={\textstyle{1\over \sqrt{2}}}\,\alpha^\dagger \cdot D_\mu(2008)
- {\textstyle{1\over \sqrt{2}}}\,D^t_\mu(2008)\cdot \alpha   +  i\,\tau_2\,D^{(s)}_{\mu}(2112) \,,
\nonumber\\
&& V_\mu^{[1]} = (J/\Psi)_\mu(3097)\,.
\label{}
\end{eqnarray}
We construct first the most general SU(3) symmetric interaction terms involving the
pseudoscalar fields:
\begin{eqnarray}
{\mathcal L}_{\rm{int}}^{\rm{SU(3)}} &=&
{\textstyle{i\over 4}}\, h_{\bar 3 \bar 3}^0\
{\rm tr}\left(
(\partial_{\mu}\Phi_{[\bar 3]})\, \Phi^\dagger_{[\bar 3]} V_{[1]}^{\mu}
-\Phi_{[\bar 3]}\,(\partial_{\mu}\Phi^\dagger_{[\bar 3]})\, V_{[1]}^{\mu}  \right)
\nonumber\\
&+& {\textstyle{i\over 4}}\, h_{9\bar 3}^{\bar 3}\
{\rm tr}\left(
(\partial_{\mu} \Phi^\dagger_{[\bar 3]})\, \Phi_{[9]}\, V_{[\bar 3]}^{\mu}-
\Phi_{[9]}\,(\partial_{\mu}\Phi_{[\bar 3]})\, V_{[\bar 3]}^{\dagger \mu}  \right)
\nonumber\\
&+& {\textstyle{i\over 4}}\, h_{\bar 39}^{\bar 3}\
{\rm tr}\left(
(\partial_{\mu}\Phi_{[9]})\, \Phi_{[\bar 3]} V_{[\bar 3]}^{\dagger \mu}-
\Phi^\dagger_{[\bar 3]}\,(\partial_{\mu}\Phi_{[9]})\, V_{[\bar 3]}^{\mu}  \right)
\nonumber\\
&+& {\textstyle{i\over 4}}\, h_{\bar 3 0}^{\bar 3}\
{\rm tr}\left(
(\partial_{\mu}\Phi_{[1]})\, \Phi_{[\bar 3]} V_{[\bar 3]}^{\dagger \mu} -
 \Phi^\dagger_{[\bar 3]}\,(\partial_{\mu}\Phi_{[1]})\, V_{[\bar 3]}^{ \mu} \right)
\nonumber\\
&+& {\textstyle{i\over 4}}\, h_{0\bar 3}^{\bar 3}\
{\rm tr}\left(
(\partial_{\mu}\Phi^\dagger_{[\bar 3]}) \,\Phi_{[1]} V_{[\bar 3]}^{\mu}
-\Phi_{[1]}\,(\partial_{\mu}\Phi_{[\bar 3]})\, V_{[\bar 3]}^{\dagger \mu}  \right)
\nonumber\\
&+& {\textstyle{i\over 4}}\, h_{\bar 3 1}^{\bar 3}\
{\rm tr}\left( \Phi^\dagger _{[\bar 3]}\,V_{[\bar 3]}^{\mu}-
\Phi_{[\bar 3]}\,V_{[\bar 3]}^{\dagger \mu}
\right)
{\tr }(\partial_{\mu}\Phi_{[9 ]})\,
\nonumber\\
&+& {\textstyle{i\over 4}}\, h_{1 \bar 3 }^{\bar 3}\
{\rm tr}\left( (\partial_{\mu} \Phi_{[\bar 3]})\,V_{[\bar 3]}^{\dagger\mu}-
(\partial_{\mu} \Phi^\dagger_{[\bar 3]})\,V_{[\bar 3]}^{ \mu}
\right)
{\tr }(\Phi_{[9 ]})\,,
\label{vector-heavy-meson}
\end{eqnarray}
where the SU(4) symmetric interaction (\ref{mesons-SU4}) suggests the identification
\begin{eqnarray}
&&h^0_{\bar 3 \bar 3} = \sqrt{2}\,g\,, \quad
h^{\bar 3}_{9 \bar 3} = 2\,g\,, \quad \;\;\,\,
h^{\bar 3}_{\bar 3 9} = 2\,g\,,
\nonumber\\
&&h^{\bar 3}_{\bar 3 0} = \sqrt{2}\,g\,, \quad
h^{\bar 3}_{0 \bar 3} = \sqrt{2}\,g\,, \quad
h^{\bar 3}_{\bar 3 1} = g\,, \quad \;
h^{\bar 3}_{1 \bar 3} = g \,.
\label{meson-SU4-result}
\end{eqnarray}
The prediction of the vertex (\ref{mesons-SU4}) can be tested against the decay pattern of
the D-meson. The parameter combination
$h^{\bar 3}_{\bar 3 9}+h^{\bar 3}_{9\bar 3 }$ determines the decay process
$D_+(2010)\to \pi_+(139)\,D_0(1865)$ as follows
\begin{eqnarray}
&&\Gamma_{D_+(2010)\to \pi_+(140)\,D_0(1865)} =
\left(\frac{h^{\bar 3}_{\bar 3 9}+h^{\bar 3}_{9\bar 3 }}{4}\right)^2\,
\frac{q_{\rm cm}^3}{8\,\pi\,M^2_{D_+(2010)}} = (64 \pm 15)\, {\rm keV} \,,
\nonumber\\
&& q_{\rm cm}^2 = \left(\frac{M^2_{D_+(2010)}+m_{\pi_+(140)}^2-M^2_{D_0(1865)}}{2\,M_{D_+(2010)}}\right)^2-m_{\pi_+(140)}^2\,.
\label{decay-d-meson}
\end{eqnarray}
From the empirical branching ratio \cite{PDG04} as recalled in
(\ref{decay-d-meson})  we deduce $(h^{\bar 3}_{\bar 3 9}+h^{\bar 3}_{9\bar 3 })/4 = 10.4 \pm 1.4$,
which is confronted with the SU(4) estimate
$(h^{\bar 3}_{\bar 3 9}+h^{\bar 3}_{9\bar 3 })/4=g \simeq 6.6 $.
We observe a small SU(4) breaking pattern. Based on this result one may
expect the relations (\ref{meson-SU4-result}) to provide magnitudes for the coupling constants
reliable within a factor two.

We close this section with the construction of the most general SU(3) symmetric interaction
involving baryon fields
\begin{eqnarray}
{\mathcal L}_{\rm{int}}^{\rm{SU(3)}} &=&
{\textstyle{1\over 2}}\ g_{88}^0 \ {\rm tr} \left( \bar B_{[8]}  \,\gamma_{\mu}\,  B_{[8]}  \,V_{[1]}^{\mu}  \right)
+ {\textstyle{1\over 2}}\ g_{66}^0  \ {\rm tr}\left( \bar B_{[6]}  \,\gamma_{\mu}\,  B_{[6]}  \,V_{[1]}^{\mu}  \right)
\nonumber\\
&+&{\textstyle{1\over 2}}\ g_{\bar 3\bar 3}^0 \ {\rm tr}\left( \bar B_{[\bar 3]}  \,\gamma_{\mu}\,  B_{[\bar 3]}  \,V_{[1]}^{\mu}  \right)
  + {\textstyle{1\over 2}}\ g_{33}^0        \ {\rm tr}\left( \bar B_{[3]}  \,\gamma_{\mu}\,  B_{[3]}  \,V_{[1]}^{\mu}   \right)
\nonumber\\
&+&{\textstyle{1\over 2}}\ g_{86}^{\bar 3}\   {\rm tr}\left( \bar B_{[8]} \,\gamma_{\mu}\, B_{[6]} \,
V_{[\bar 3]}^{\dagger \mu}
+  \bar B_{[6]} \,\gamma_{\mu}\, B_{[8]} \,V_{[\bar 3]}^{\mu} \right)
\nonumber\\
&+& {\textstyle{1\over 2}}\ g_{8\bar{3}}^{\bar 3}\ {\rm tr}\left( \bar B_{[8]} \,\gamma_{\mu}\, B_{[\bar 3]}
\,V_{[\bar 3]}^{\dagger \mu}  +
 \bar B_{[\bar 3]} \,\gamma_{\mu}\, B_{[8]} \,V_{[\bar 3]}^{\mu} \right)
\nonumber\\
&+&{\textstyle{1\over 2}}\ g_{63}^{\bar 3}\
{\rm{tr}}\ \Big( \bar B_{[6]}\,\gamma_{\mu} \,
{\rm{tr}} \big( B_{[3]}\,V_{[\bar 3]}^{\dagger\mu} \big)
-2\, \bar B_{[6]}\,\gamma_{\mu}\,B_{[3]}\,V_{[\bar 3]}^{\dagger\mu}
\nonumber\\
& & \qquad \;\;\;\;+ \bar B_{[3]}\,\gamma_{\mu}\,V_{[\bar 3]}^{\phantom{\dagger}\mu}
\,{\rm{tr}}\, \big( B_{[6]} \big)
-2\,\bar B_{[3]}\,\gamma_{\mu}\, B_{[6]}\,V_{[\bar 3]}^{\phantom{\dagger}\mu}
\Big)
\nonumber\\
&+&{\textstyle{1\over 2}}\ g_{\bar 33}^{\bar 3}\
{\rm{tr}} \left( \bar B_{[\bar 3]}\,\gamma_{\mu}\,B_{[3]}\,V_{[\bar 3]}^{\dagger\mu}
+ \bar B_{[3]}\,\gamma_{\mu}\,B_{[\bar 3]}\,V_{[\bar 3]}^{\phantom{\dagger}\mu} \right) \,,
\label{vector-heavy-baryon}
\end{eqnarray}
where the SU(4) symmetric vertex (\ref{baryons-SU4}) implies
\begin{eqnarray}
&& g^0_{88 } = 0\,, \qquad
g^{0}_{66} = \sqrt{2}\,g\,, \qquad
g^{0}_{\bar 3 \bar 3} =\sqrt{2}\, g\,, \qquad
g^{0}_{3  3} = 2\,\sqrt{2}\,g\,, \qquad
\nonumber\\
&& g^{\bar 3}_{86} =\sqrt{2}\, g\,, \qquad
g^{\bar 3}_{8 \bar 3} =-\sqrt{6}\, g\,, \qquad
g^{\bar 3}_{6 3} =g\, ,\qquad
g^{\bar 3}_{\bar 3 3} = 2\,\sqrt{3}\,g\,. \qquad
\label{baryon-SU4-result}
\end{eqnarray}
Unfortunately there appears to be no way at present to check on the usefulness
of the result (\ref{baryon-SU4-result}). Eventually simulations of QCD on a lattice may
shed some light on this issue. The precise values of the coupling constants
(\ref{meson-SU4-result}, \ref{baryon-SU4-result})
will not affect the major results of this work. This holds as long as those coupling constants
range in the region suggested by (\ref{meson-SU4-result}, \ref{baryon-SU4-result}) within a
factor two.

\section{Coupled-channel scattering}

\begin{table}[t]\rescale
\setlength{\tabcolsep}{1.4mm}
\setlength{\arraycolsep}{1.65mm}
\renewcommand{\arraystretch}{1.5}
\begin{center}
\tabcolsep=2mm
\begin{tabular}{|@{}c@{}c@{}c@{}c@{}c@{}c@{}c@{}c@{}c@{}c@{}c@{}c@{}|}
\hline \hline
\multicolumn{3}{|c|}{\makebox[3cm][c]{$(0,0,-1)$}} &
\multicolumn{3}{c|}{\makebox[3cm][c]{$(1,0,-1)$}} &
\multicolumn{3}{c|}{\makebox[3cm][c]{$(\frac12,-1,-1)$}} &
\multicolumn{3}{c|}{\makebox[3cm][c]{$(\frac32,-1,-1)$}}\\
\hline
\multicolumn{3}{|c|}{$(\frac{1}{\sqrt2}\bar D^t\,i\sigma_2 N) $} &
\multicolumn{3}{c|}{$ ( \frac{1}{\sqrt2}\bar D^t\,i\sigma_2\vec{\sigma}\, N ) $}&
\multicolumn{3}{c|}{$ \left(\begin{array}{c} (\rule{0pt}{2.3ex} \bar D_s \,N) \\ (\Lambda\, \bar D) \\ (\frac1{\sqrt3}\Sigma \cdot \sigma \bar{D}) \end{array}\right) $} &
\multicolumn{3}{c|}{$ (\Sigma \cdot T \,\bar D) $}\\
\hline \hline
\multicolumn{4}{|c|}{\makebox[4.15cm][c]{$(0,-2,-1)$}} &
\multicolumn{4}{c|}{\makebox[4.15cm][c]{$(1,-2,-1)$}} &
\multicolumn{4}{c|}{\makebox[4.15cm][c]{$(\frac12,-3,-1)$}}\\
\hline
\multicolumn{4}{|c|}{$\left(\begin{array}{c} (\bar{D}_s\, \Lambda) \\
(\frac{1}{\sqrt2} \bar{D}^t\, i \sigma_2\, \Xi) \end{array}\right) $} &
\multicolumn{4}{c|}{$\left(\begin{array}{c}  (\bar{D}_s \,\vec{\Sigma}) \\
(\frac{1}{\sqrt2} \bar{D}^t \,i \sigma_2 \vec{\sigma}\, \Xi) \end{array}\right) $} &
\multicolumn{4}{c|}{$ (\bar{D}_s \,\Xi) $}\\
\hline \hline
\multicolumn{3}{|c|}{\makebox[3cm][c]{$(0,1,0)$}} &
\multicolumn{3}{c|}{\makebox[3cm][c]{$(1,1,0)$}} &
\multicolumn{3}{c|}{\makebox[3cm][c]{$(\frac12,0,0)$}} &
\multicolumn{3}{c|}{\makebox[3cm][c]{$(\frac32,0,0)$}}\\
\hline
\multicolumn{3}{|c|}{$ (\frac{1}{\sqrt2} K^t\, i \sigma_2\, N) $} &
\multicolumn{3}{c|}{$ (\frac{1}{\sqrt2} K^t \,i \sigma_2 \vec{\sigma}\, N) $} &
\multicolumn{3}{c|}{$\left(\begin{array}{c} (\frac{1}{\sqrt3} \pi \cdot \sigma \,N) \\
(\eta \,N) \\
(\Lambda\, K) \\
(\frac{1}{\sqrt3} \Sigma \cdot \sigma \,K) \\
(\eta' \,N) \\
(\eta_c \,N) \\
(\Lambda_c\, \bar{D}) \\
(\frac{1}{\sqrt3} \Sigma_c \cdot \sigma \,\bar{D})
\end{array}\right) $} &
\multicolumn{3}{c|}{$ \left(\begin{array}{c} (\pi \cdot T \,N) \\
(\Sigma \cdot T \,K) \\ (\Sigma_c \cdot T \,\bar{D})
\end{array}\right) $}\\
\hline \hline
\multicolumn{3}{|c|}{$(0,-1,0)$} &
\multicolumn{3}{c|}{$(1,-1,0)$} &
\multicolumn{3}{c|}{$(2,-1,0)$} &
\multicolumn{3}{c|}{$(\frac12,-2,0)$}\\
\hline
\multicolumn{3}{|c|}{$\left(\begin{array}{c} (\frac{1}{\sqrt3}\Sigma \cdot \pi) \\
(\frac{1}{\sqrt2} \bar{K} N) \\
(\Lambda \,\eta) \\
(\frac{1}{\sqrt2} K^t\, i \sigma_2 \,\Xi) \\
(\Lambda \,\eta') \\
(\Lambda \,\eta_c) \\
(\Lambda_c \,\bar{D}_s) \\
(\frac{1}{\sqrt2} \bar{D}^t\, i \sigma_2\, \Xi_c) \\
(\frac{1}{\sqrt2} \bar{D}^t \,i \sigma_2 \,\Xi_c')
\end{array}\right) $} &
\multicolumn{3}{c|}{$\left(\begin{array}{c} (\Lambda \,\vec{\pi}) \\
(\frac{i}{\sqrt2} \vec{\Sigma} \times \vec{\pi}) \\
(\frac{1}{\sqrt2} \bar{K} \vec{\sigma}\, N) \\
(\eta \,\vec{\Sigma}) \\
(\frac{1}{\sqrt2} K^t\, i \sigma_2 \vec{\sigma}\, \Xi) \\
(\eta' \vec{\Sigma}) \\
(\eta_c \vec{\Sigma}) \\
(\frac{1}{\sqrt2} \bar{D}^t\, i \sigma_2 \vec{\sigma}\, \Xi_c) \\
(\bar{D}_s \vec{\Sigma}_c) \\
(\frac{1}{\sqrt2} \bar{D}^t\, i \sigma_2 \vec{\sigma}\, \Xi_c')
\end{array}\right) $} &
\multicolumn{3}{|c|}{$ (\Sigma \cdot S \cdot \pi) $} &
\multicolumn{3}{c|}{$\left(\begin{array}{c}
(\frac{1}{\sqrt3} \pi \cdot \sigma \,\Xi) \\
(\Lambda \,i \sigma_2 \,\bar{K}^t) \\
(\frac{1}{\sqrt3} \Sigma\, \cdot \sigma i \sigma_2\, \bar{K}^t) \\
(\eta\, \Xi) \\
(\eta'\, \Xi) \\
(\eta_c\, \Xi) \\
(\bar{D}_s \,\Xi_c) \\
(\bar{D}_s \,\Xi_c') \\
(\Omega_c \,\bar{D})
\end{array}\right) $}\\
\hline \hline
\multicolumn{4}{|c|}{\makebox[4.15cm][c]{$(\frac32,-2,0)$}} &
\multicolumn{4}{c|}{\makebox[4.15cm][c]{$(0,-3,0)$}} &
\multicolumn{4}{c|}{\makebox[4.15cm][c]{$(1,-3,0)$}}\\
\hline
\multicolumn{4}{|c|}{$\left(\begin{array}{c}
(\pi \cdot T\, \Xi) \\
(\Sigma \cdot T \,i \sigma_2 \,\bar{K}^t)
\end{array}\right) $} &
\multicolumn{4}{|c|}{$\rule{0pt}{4ex} \left(\begin{array}{c}
(\frac{1}{\sqrt2}  \bar{K} \,\Xi) \\
(\bar{D}_s \,\Omega_c)
\end{array}\right) $} &
\multicolumn{4}{|c|}{$
(\frac{1}{\sqrt2} \bar{K}\, \vec{\sigma}\, \Xi) $}\\
\hline
\end{tabular}
\caption{Definition of coupled-channel states with isospin ($I$), strangeness ($S$) and
charm ($C$). Here $C=-1,0$.}
\label{tab:states1}
\end{center}
\end{table}

\begin{table}[t]\rescale
\setlength{\tabcolsep}{1.4mm}
\setlength{\arraycolsep}{2.8mm}
\renewcommand{\arraystretch}{1.37}
\begin{tabular}{|cccccccccccc|}
\hline
\hline
\multicolumn{3}{|c|}{\phantom{xxxx}$(\frac12,1,1)$\phantom{xxxx}} &
\multicolumn{3}{c|}{\phantom{xxxx}$(\frac32,1,1)$\phantom{xxxx}} &
\multicolumn{3}{c|}{\phantom{xxxx}$(0,0,1)$\phantom{xxxx}} &
\multicolumn{3}{c|}{\phantom{xxxx}$(1,0,1)$\phantom{xxxx}}\\
\hline
\multicolumn{3}{|c|}{$\left(
\begin{array}{c}
(\Lambda_c \,K) \\
(D_s \,N) \\
(\frac{1}{\sqrt3} \Sigma_c \cdot \sigma \,K)
\end{array}\right) $} &
\multicolumn{3}{c|}{$ (\Sigma_c \cdot T \,K) $} &
\multicolumn{3}{c|}{$\left(
\begin{array}{c} (\frac{1}{\sqrt3} \pi \cdot \Sigma_c) \\
(\frac{1}{\sqrt2} D \,N) \\
(\Lambda_c \,\eta) \\
(\frac{1}{\sqrt2} K^t\, i \sigma_2 \,\Xi_c) \\
(\frac{1}{\sqrt2} K^t\, i \sigma_2 \,\Xi_c') \\
(D_s \,\Lambda) \\
(\Lambda_c \,\eta') \\
(\Lambda_c \,\eta_c) \\
(\frac{1}{\sqrt2} \bar{D}^t\, i \sigma_2 \,\Xi_{cc})
\end{array}\right) $} &
\multicolumn{3}{c|}{$\left(
\begin{array}{c}  (\Lambda_c \,\vec{\pi}) \\
(\frac{i}{\sqrt2} \vec{\Sigma}_c \times \vec{\pi}) \\
(\frac{1}{\sqrt2} D\, \vec{\sigma}\, N) \\
(\frac{1}{\sqrt2} K^t \,i \sigma_2 \vec{\sigma} \,\Xi_c) \\
(\eta \,\vec{\Sigma}_c) \\
(\frac{1}{\sqrt2} K^t\, i \sigma_2 \vec{\sigma} \,\Xi_c') \\
(D_s \,\vec{\Sigma}) \\
(\eta' \,\vec{\Sigma}_c) \\
(\frac{1}{\sqrt2} \bar{D}^t\, i \sigma_2 \vec{\sigma}\, \Xi_{cc}) \\
(\eta_c \,\vec{\Sigma}_c)
\end{array}\right) $}\\
\hline \hline
\multicolumn{3}{|c|}{$(2,0,1)$} &
\multicolumn{3}{c|}{$(\frac12,-1,1)$} &
\multicolumn{3}{c|}{$(\frac32,-1,1)$} &
\multicolumn{3}{c|}{$(0,-2,1)$}\\
\hline
\multicolumn{3}{|c|}{$ (\Sigma_c \cdot S \cdot \pi) $} &
\multicolumn{3}{c|}{$\left(
\begin{array}{c}
(\frac{1}{\sqrt3} \pi \cdot \sigma \,\Xi_c) \\
(\frac{1}{\sqrt3} \pi \cdot \sigma \,\Xi_c') \\
(\Lambda_c\, i \sigma_2 \,\bar{K}^t) \\
(\frac{1}{\sqrt3} \Sigma_c \cdot \sigma i \sigma_2 \,\bar{K}^t) \\
(\Lambda\, i \sigma_2\, D^t) \\
(\eta\, \Xi_c) \\
(\frac{1}{\sqrt3} \Sigma \cdot \sigma i \sigma_2 \,D^t) \\
(\eta \,\Xi_c') \\
(\Omega_c \,K) \\
(D_s \,\Xi) \\
(\eta'\, \Xi_c) \\
(\eta'\, \Xi_c') \\
(\eta_c \,\Xi_c) \\
(\bar D_s \,\Xi_{cc}) \\
(\Omega_{cc}\, \bar{D}) \\
(\eta_c \,\Xi_c')
\end{array}\right) $} &
\multicolumn{3}{c|}{$\left(\begin{array}{c}
(\pi \cdot T \,\Xi_c) \\
(\pi \cdot T \,\Xi_c') \\
(\Sigma_c \cdot T \,i \sigma_2 \,\bar{K}^t) \\
(\Sigma \cdot T i \sigma_2 D^t)
\end{array}\right) $} &
\multicolumn{3}{c|}{$\left(
\begin{array}{c} (\frac{1}{\sqrt2} \bar{K} \,\Xi_c) \\
(\frac{1}{\sqrt2} \bar{K}\, \Xi_c') \\
(\frac{1}{\sqrt2} D \,\Xi) \\
(\Omega_c \,\eta) \\
(\Omega_c \,\eta') \\
(\Omega_{cc}\, \bar D_s) \\
(\Omega_c \,\eta_c)
 \end{array}\right) $}\\
\hline \hline
\multicolumn{6}{|c|}{$(1,-2,1)$} &
\multicolumn{6}{c|}{$(\frac12,-3,1)$}\\
\hline
\multicolumn{6}{|c|}{$\left(
\begin{array}{c} (\Omega_c \,\vec{\pi}) \\
(\frac{1}{\sqrt2} \bar{K} \,\vec{\sigma} \,\Xi_c) \\
(\frac{1}{\sqrt2} \bar{K}\, \vec{\sigma} \,\Xi_c') \\
(\frac{1}{\sqrt2} D \,\vec{\sigma} \,\Xi)
\end{array}\right) $} &
\multicolumn{6}{c|}{$ (\Omega_c\, i \sigma_2 \,\bar{K}^t) $}\\
\hline
\hline
\end{tabular}
\caption{Definition of coupled-channel states with $C=1$.}
\label{tab:states2}
\end{table}

\begin{table}\rescale
\setlength{\tabcolsep}{1.4mm}
\setlength{\arraycolsep}{2.7mm}
\renewcommand{\arraystretch}{1.4}
\begin{tabular}{|cccccccccccc|}
\hline
\hline
\multicolumn{3}{|c|}{\makebox[3cm][c]{$(0,1,2)$}} &
\multicolumn{3}{c|}{\makebox[3cm][c]{$(1,1,2)$}} &
\multicolumn{3}{c|}{\makebox[3cm][c]{$(\frac12,0,2)$}} &
\multicolumn{3}{c|}{\makebox[3cm][c]{$(\frac32,0,2)$}}\\
\hline
\multicolumn{3}{|c|}{$\left(
\begin{array}{c}
\makebox[1.6cm][c]{$(\frac1{\sqrt2}K^t \,i \sigma_2\, \Xi_{cc})$} \\
(\Lambda_c \,D_s)
\end{array}\right) $} &
\multicolumn{3}{c|}{$\left(
\begin{array}{c}
\makebox[1.6cm][c]{$(\frac1{\sqrt2}K^t\, i \sigma_2 \vec{\sigma}\, \Xi_{cc})$} \\
(D_s \,\vec{\Sigma}_c)
\end{array}\right) $} &
\multicolumn{3}{c|}{$\left(
\begin{array}{c} (\frac1{\sqrt3} \pi \cdot \sigma \Xi_{cc}) \\
(\eta \,\Xi_{cc}) \\
(\Omega_{cc} \,K) \\
(\Lambda_c\, i \sigma_2\, D^t) \\
(\frac1{\sqrt3} \Sigma_c \cdot \sigma i \sigma_2\, D^t) \\
(\eta' \,\Xi_{cc}) \\
(D_s \,\Xi_c) \\
(D_s \,\Xi_c') \\
(\eta_c \,\Xi_{cc})
\end{array}\right) $} &
\multicolumn{3}{c|}{$\left(
\begin{array}{c} (\pi \cdot T \,\Xi_{cc}) \\
(\Sigma_c \cdot T \,i \sigma_2 \,D^t)
\end{array}\right) $}\\
\hline \hline
\multicolumn{4}{|c|}{\makebox[4.15cm][c]{$(0,-1,2)$}} &
\multicolumn{4}{c|}{\makebox[4.15cm][c]{$(1,-1,2)$}} &
\multicolumn{4}{c|}{\makebox[4.15cm][c]{$(\frac12,-2,2)$}} \\
\hline
\multicolumn{4}{|c|}{$\left(
\begin{array}{c} (\rule{0pt}{2.3ex}
\frac1{\sqrt2} \bar K \,\Xi_{cc}) \\
(\Omega_{cc} \,\eta) \\
(\frac1{\sqrt2} D \,\Xi_c) \\
(\frac1{\sqrt2} D \,\Xi_c') \\
(\Omega_{cc} \,\eta') \\
(D_s \,\Omega_c) \\
(\Omega_{cc} \,\eta_c)
\end{array}\right) $} &
\multicolumn{4}{c|}{$\left(
\begin{array}{c}
(\Omega_{cc}\, \vec{\pi}) \\
(\frac1{\sqrt2} \bar K\, \vec{\sigma} \,\Xi_{cc}) \\
(\frac1{\sqrt2} D \,\vec{\sigma} \,\Xi_c) \\
(\frac1{\sqrt2} D \,\vec{\sigma} \,\Xi_c')
\end{array}\right) $} &
\multicolumn{4}{c|}{$\left(
\begin{array}{c}
(\Omega_{cc} \,i \sigma_2 \,\bar K^t) \\
(\Omega_{c} \,i \sigma_2 \,D^t)
\end{array}\right) $} \\
\hline
\hline
\multicolumn{3}{|c|}{\makebox[3cm][c]{$(\frac12,1,3)$}} &
\multicolumn{3}{c|}{\makebox[3cm][c]{$(0,0,3)$}} &
\multicolumn{3}{c|}{\makebox[3cm][c]{$(1,0,3)$}} &
\multicolumn{3}{c|}{\makebox[3cm][c]{$(\frac12,-1,3)$}}\\
\hline
\multicolumn{3}{|c|}{$ (D_s\, \Xi_{cc}) $} &
\multicolumn{3}{c|}{$\left(
\begin{array}{c}
(\frac1{\sqrt2} D \,\Xi_{cc}) \\
(\Omega_{cc}\, D_s)
\end{array}\right) $} &
\multicolumn{3}{c|}{$ (\frac1{\sqrt2} D \,\vec{\sigma}\, \Xi_{cc}) $} &
\multicolumn{3}{c|}{$ (\Omega_{cc}\, i \sigma_2 \,D^t) $} \\
\hline
\hline
\end{tabular}
\caption{Definition of coupled-channel states with $C=2,3$.}
\label{tab:states3}
\end{table}

We consider the s-wave scattering of the pseudo-scalar mesons fields (\ref{mesons})
off the baryon fields (\ref{baryons}). The scattering kernel is approximated
by the t-channel vector meson exchange force defined by
(\ref{vector-pseudoscalar}, \ref{vector-baryons}), where we
apply the formalism developed in \cite{LK02,LK04-axial}. The scattering kernel has
the form
\begin{eqnarray}
K^{(I,S,C)}(\bar q,q;w) = -\frac{1}{4}\,\sum_{V \in [16]}\,
\frac{C^{(I,S,C)}_V}{t-m^2_V} \,\Big( \frac{\barqslash +\qslash }{2}
- (\bar q^2-q^2)\,\frac{\barqslash-\qslash}{2\,m_V^2}\Big)\,,
\label{def-K}
\end{eqnarray}
with the initial and final meson 4-momenta $q_\mu$ and $\bar q_\mu$ and $t=(\bar q-q)^2$.
It should be mentioned that (\ref{def-K}) is valid only under the assumptions
$h_{\bar 39}^{\bar 3}=h_{9\bar 3}^{\bar 3}$, $h_{\bar 31}^{\bar 3}=h_{1\bar 3}^{\bar 3}$
and $h_{\bar 30}^{\bar 3}=h_{0\bar 3}^{\bar 3}$.
In (\ref{def-K}) the scattering is projected onto sectors with conserved isospin (I),
strangeness (S) and
charm (C) quantum numbers. The latter are introduced with respect to the states collected in
Tabs. \ref{tab:states1}-\ref{tab:states3}, where we use the notation and conventions of
\cite{LK02,LWF02}. The coupled-channel structure of the matrices $C_{V,ab}^{(I,S,C)}$ is
given in the Appendix. Only non-vanishing elements are displayed.
Owing to the 'chiral' identifications
(\ref{chiral-constraint-mesons}, \ref{chiral-constraint-baryons}) and the KSFR relation
(\ref{KSFR-relation}) we reproduce the coupled-channel structure of the Weinberg-Tomozawa
interaction identically. Summing over the light vector meson states
\begin{eqnarray}
\sum_{V\in [9]}\,C^{(I,S,C)}_V =4\,g^2\,C_{WT}^{(I,S,C)} \,,
\label{WT-repr}
\end{eqnarray}
we reproduce the matrices $C_{WT}$ as given previously in \cite{LK02,LK04-charm}.
The first term of the interaction kernel matches corresponding expressions predicted
by the leading order chiral Lagrangian if we put $t=0$ in (\ref{def-K}) and use the common
value for the vector-meson masses suggested by the KSFR relation (\ref{KSFR-relation}).
The second term in (\ref{def-K}) is formally of chiral order $Q^3$ for
channels involving Goldstone bosons. Numerically it is a minor correction but nevertheless it is
kept in the computation.

Given (\ref{meson-SU4-result}, \ref{baryon-SU4-result}) one
may decompose the interaction into SU(4) invariant tensors:
\begin{eqnarray}
&&\frac{1}{4\,g^2}\sum_{V\in[16]}\,C^{(I,S,C)}_V =
7\,C_{[\overline{4}]}+4\,C_{[20_1]} +4\,C_{[20_2]}+ C_{[20_s]}
+2\,C_{[\overline{36}]}-2\,C_{[140]}  \,,
\nonumber\\ \nonumber\\
&& \qquad \qquad 15 \otimes 20 =  \overline{4}
\oplus 20_1 \oplus 20_2 \oplus 20_s  \oplus \overline{36}\oplus \overline{60}  \oplus 140 \,.
\label{SU4-decomposition}
\end{eqnarray}
The normalization of the matrices $C_{[...]}$ is such that their weight factors in
(\ref{SU4-decomposition}) give the eigenvalues
of the coupling matrix $C$. Strongest attraction is foreseen in the $\bar 4$-plet,
repulsion in the 140-plet and no interaction in the 60-plet.
Three distinct 20-plet are formed, all receiving identical attractive weight factors.
It is interesting to observe that (\ref{SU4-decomposition}) predicts attraction in
the $\overline{36}$-plet. This is an exotic multiplet. In order to help the reader to digest this
abstract group theory we further decompose the SU(4) multiplets
into the more familiar SU(3) multiplets
\begin{eqnarray}
&&\;\, [\overline 4]^{\rm{SU(4)}} =\;\,[1]^{\rm{SU(3)}}_{C=0\,} \oplus [\overline{3}]^{\rm{SU(3)}}_{C=1} \,,
\nonumber\\
&&[20 ]_{1,2}^{\rm{SU(4)}} = \;\,[8]^{\rm{SU(3)}}_{C=0\,} \oplus [6]^{\rm{SU(3)}}_{C=1} \oplus
[\overline{3}]^{\rm{SU(3)}}_{C=1}
\oplus [3]^{\rm{SU(3)}}_{C=2}  \,,
\nonumber\\
&&[20 ]_{s}^{\rm{SU(4)}} = [10]^{\rm{SU(3)}}_{C=0} \oplus [6]^{\rm{SU(3)}}_{C=1}
\oplus [3]^{\rm{SU(3)}}_{C=2} \oplus
[1]^{\rm{SU(3)}}_{C=3} \,,
\nonumber\\
&&[\overline{36} ]^{\rm{SU(4)}} = \;\,[3]^{\rm{SU(3)}}_{C=-1} \oplus [8]^{\rm{SU(3)}}_{C=0} \oplus
[\overline{15}]^{\rm{SU(3)}}_{C=1}
\oplus [\overline{3}]^{\rm{SU(3)}}_{C=1} \oplus [\overline{6}]^{\rm{SU(3)}}_{C=2} \,.
\label{36-decomp}
\end{eqnarray}
From (\ref{SU4-decomposition}, \ref{36-decomp}) one may expect the formation of penta-quark type
states with negative charm. The persistence of attraction in the exotic $\overline{15}$-plet sector
was claimed before in \cite{LK04-charm}. It should be stressed that the decomposition
(\ref{SU4-decomposition}) is more of academic interest, even though it is useful to
perform some consistency checks. One carefully has to explore whether
the attraction is provided by the t-channel exchange of the light or heavy vector mesons.
This important piece of information is lost in (\ref{SU4-decomposition}).

In this work we neglect the t-dependence of the interaction kernel
insisting on $t=0$ in (\ref{def-K}). Following \cite{LK02,LK04-axial} the s-wave
projected effective scattering kernel, $V^{(I,S,C)}(\sqrt{s}\,)$, is
readily constructed:
\begin{eqnarray}
V^{(I,S,C)}(\sqrt{s}\,) = \!\!\sum_{V\in [16]}\frac{C^{(I,S,C)}_V}{8\,m_V^2}\, \Big(
2\,\sqrt{s}-M-\bar M +(\bar M-M)\,\frac{\bar m^2 -m^2}{m_V^2} \Big) \,,
\label{VWT}
\end{eqnarray}
where $M$, $\bar M$ and $m, \bar m$ are the masses of initial and final
baryon and meson states. The scattering
amplitudes, $M^{(I,S,C)}(\sqrt{s}\,)$,  take the simple form
\begin{eqnarray}
&&  M^{(I,S,C)}(\sqrt{s}\,) = \Big[ 1- V^{(I,S,C)}(\sqrt{s}\,)\,J^{(I,S,C)}(\sqrt{s}\,)\Big]^{-1}\,
V^{(I,S,C)}(\sqrt{s}\,)\,.
\label{final-t}
\end{eqnarray}
The unitarity loop function, $J^{(I,S,C)}(\sqrt{s}\,)$, is a diagonal matrix. Each element
depends on the masses of intermediate meson and baryon, $m$ and $M$, respectively:
\begin{eqnarray}
&& J(\sqrt{s}\,) =
\Big(M + (M^2+p_{\rm cm}^2)^{1/2} \Big)\,
\Big(I(\sqrt{s}\,)-I(\mu) \Big)\,,
\nonumber\\
\nonumber\\
&& I(\sqrt{s}\,)=\frac{1}{16\,\pi^2}
\left( \frac{p_{\rm cm}}{\sqrt{s}}\,
\left( \ln \left(1-\frac{s-2\,p_{\rm cm}\,\sqrt{s}}{m^2+M^2} \right)
-\ln \left(1-\frac{s+2\,p_{\rm cm}\sqrt{s}}{m^2+M^2} \right)\right)
\right.
\nonumber\\
&&\qquad \qquad + \left.
\left(\frac{1}{2}\,\frac{m^2+M^2}{m^2-M^2}
-\frac{m^2-M^2}{2\,s}
\right)
\,\ln \left( \frac{m^2}{M^2}\right) +1 \right)+I(0)\;,
\label{i-def}
\end{eqnarray}
where $\sqrt{s}= \sqrt{M^2+p_{\rm cm}^2}+ \sqrt{m^2+p_{\rm cm}^2}$.
A crucial ingredient of the approach developed in \cite{LK02,LK04-axial} is its
approximate crossing symmetry guaranteed by a proper choice of the subtraction scales $\mu $.
The latter depends on the quantum number $(I,S,C)$ but should be chosen uniformly
within a given sector \cite{LK02,LK04-axial,Granada,Copenhagen}.
We insist on
\begin{eqnarray}
&& \mu =
\sqrt{  m_{\rm th}^2+M_{\rm th}^2}\,, \qquad
m_{\rm th}+M_{\rm th} = {\rm Min} \{m_a+M_a \} \,.
\label{mu-def}
\end{eqnarray}
As a consequence of (\ref{mu-def}) the s-channel and u-channel unitarized amplitudes
involving the lightest channels can be matched smoothly at the subtraction point
$\mu$ \cite{LK02,LK04-axial,Granada,Copenhagen}.
The construction (\ref{mu-def}) implies that the effect of heavy channels, like the
$\eta_c \,N$ channel, on the light channels, like the $\pi \,N$ channel, is naturally
suppressed. Since the $\eta_c\,N$ loop function is extremely smooth for
$\sqrt{s} \ll m_{\eta_c}+M_N$, enforcing the loop to vanish at $s=\mu^2 = m_\pi^2+m_N^2$
ensures that the heavy channel has a negligible effect on the low-energy scattering
of the light channel. Such a mechanism is required as to prevent an uncontrolled
renormalization of the Weinberg-Tomozawa interaction strength. If the $\eta_c\,N$ loop
would have a significant size at  $s=m_N^2+m_\pi^2$, integrating out the $\eta_c$-field
would predict an effective scattering kernel for the $\pi \,N \to \pi \,N$ system that
is in conflict with chiral symmetry.

\clearpage

\section{Numerical results}

In order to study the formation of baryon resonances we produce generalized speed plots
\cite{Hoehler:speed,LK04-axial} of the simple form
\begin{eqnarray}
{\rm Speed}_{ab}(\sqrt{s}\,) =
\Big| \frac{d}{ d \sqrt{s}\,}\,\big[\,M_{ab}(\sqrt{s}\,) \big] \Big|\,.
 \label{def-gen-speed}
\end{eqnarray}
If a partial-wave scattering amplitude develops a
resonance or bound state, close to that structure it may be approximated by
a pole and a background term. We write
\begin{eqnarray}
&& M_{ab}(\sqrt{s}\,) \simeq  -\frac{g^*_a\,g^{}_b}{\sqrt{s}-M_R+i\,\Gamma_R/2} + b_{ab} \,,
\label{Breit-Wigner}
\end{eqnarray}
with the resonance mass $M_R$ and width $\Gamma_R$. The dimension less
coupling constants $g_b$ and $g_a$  parameterize the coupling strength of
the resonance to the initial and final channels. The background term $b_{ab}$
is in general a complex number. If the scattering amplitude  has the form (\ref{Breit-Wigner})
its speed takes a maximum at the resonance mass $M_R$. The ratio of coupling constants to
total decay width $\Gamma_R$ is then determined by the value the speed takes at its maximum
\begin{eqnarray}
{\rm Speed}_{aa}(M_R) = \Bigg|\frac{2\,g_a}{\Gamma_R} \Bigg|^2 \,.
\label{def-speed}
\end{eqnarray}
We determine the resonance position and coupling constants by adjusting the parameters
$M_R$ and $g_a$ to the Speed of its associated amplitudes. This is an approximate
procedure, fully sufficient in view of the schematic nature of the computation.
We will present the Speed functions on a logarithmic scale. This presentation has the advantage
that a resonance signal can be cleanly separated from a cusp effect. Whereas a cusp effect
leads to a structure confined  in small interval around the threshold that defines the cusp,
a resonance defines a much more extended ar\^ete-like form.

If a resonance is produced in a two-body collision the formation cross section can be
expressed in terms of the coupling constants as follows
\begin{eqnarray}
&& \sum_b\sigma^{\rm form}_{a\to b} =\frac{\sqrt{M_a^2+q_{\rm cm}^2}+M_a}{q_{\rm cm}\,
\Gamma_R\,M_R}\, g_a^2\,, \quad \!\!
 M_R=\sqrt{M_a^2+q_{\rm cm}^2}+\sqrt{m_a^2+q_{\rm cm}^2}\,,
\label{formation-cross-section}
\end{eqnarray}
where it is assumed that the initial state has a well defined isospin. An appropriate Clebsch-Gordon
coefficient has to be supplied that projects the initial state onto
the isospin quantum numbers of the resonance considered.

\subsection{S-wave resonances with charm minus one}

We begin with a presentation of results obtained for $J^P=\frac{1}{2}^-$ resonances with
negative charm. The possible existence of such states was discussed by Lipkin twenty years ago
\cite{Lipkin:87}. Such states were studied in a quark model \cite{Genovese:98}, however,
predicting that they are unbound. Recently the H1 collaboration reported on a possible
signal for a state with negative charm and zero strangeness at 3.099 GeV \cite{H1Collaboration}.
This triggered various theoretical investigations
\cite{Stewart:04,Kim:04,Wessling:04,Pirjol:Schat:05,Bicudo:05}. According to a QCD sum rule
study \cite{Kim:04} the H1 signal should be associated with a p-wave resonance. Since
we consider only s-wave resonances in this work, we do not necessarily expect a signal in
this sector. Confirming the anticipation of Lipkin \cite{Lipkin:87} we observe bound states with
$(I,S)=(\frac{1}{2},-1)$ and $(0,-2)$ only. No bound or resonance-state signal are seen in
the remaining sectors.

\begin{table}[t]
\rescale
\setlength{\tabcolsep}{1.2mm}
\setlength{\arraycolsep}{2.2mm}
\renewcommand{\arraystretch}{0.75}
\begin{center}
\begin{tabular}{|ll|c|c|c|c|c|}
\hline $C=-1:$ &$ (\,I,\phantom{+}S)$  &
$\rm state$ &  $\begin{array}{c} M_R [\rm MeV]  \\ \Gamma_R \,[\rm MeV]  \end{array}$ &
$|g_R|$  & $\begin{array}{c} M_R [\rm MeV]  \\ \Gamma_R \,[\rm MeV]  \end{array}$ & $|g_R|$  \\
\hline \hline
&$(\frac12,-1)$  &
$\begin{array}{l}  \rule{0pt}{0.3ex}
\bar D_s N \\
\bar D \,\Lambda \\
\bar{D} \,\Sigma
\end{array}$
& $\begin{array}{c} 2687 \\ 0 \end{array}$ & $\begin{array}{c} 3.8 \\ 1.4 \\ 5.4 \end{array}$
& $\begin{array}{c} 2780 \\ 0 \end{array}$ & $\begin{array}{c} 3.3 \\ 1.1 \\ 4.9 \end{array}$\\
\hline
&$(0,-2)$    & $
\begin{array}{l}
\bar{D}_s \Lambda \\
\bar{D} \,\Xi
\end{array}$
& $\begin{array}{c} 2763 \\ 0  \end{array}$ & $\begin{array}{c} 3.4 \\ 6.1 \end{array}$
& $\begin{array}{c} 2838 \\ 0 \end{array}$ & $\begin{array}{c} 3.0 \\ 5.6 \end{array}$\\
\hline
\end{tabular}
\caption{Spectrum of $J^P=\frac{1}{2}^-$ baryons with charm minus one. The 3rd and 4th columns follow
with the universal vector coupling constant $g =6.6$ and SU(4) symmetric 3-point vertices. In the
5th and 6th columns we use the SU(4) breaking relation  $h_{\bar 3 \bar 3}^1  \simeq -1.19 \,g$,
which  implies a non-vanishing OZI violating $\phi_\mu \bar D D$ vertex.
The remaining SU(4) relations (\ref{meson-SU4-result}, \ref{baryon-SU4-result}) are untouched.}
\label{tab:result1}
\end{center}
\end{table}

The properties of the states are collected in Tab. \ref{tab:result1}.
The masses and coupling constants are shown for two cases. First the spectrum is computed insisting
on the chiral relations (\ref{chiral-constraint-mesons}, \ref{chiral-constraint-baryons})
together with the leading order large-$N_c$ relations (\ref{OZI-constraint}). Relying on
the KSFR relation (\ref{KSFR-relation}, \ref{chiral-constraint-mesons}) the binding energies
are determined by the universal vector coupling constant for which we take the
value $g =6.6$ in this work. The masses of the $(\frac{1}{2},-1)$ and $(0,-2)$ states are
predicted at 2.69 GeV and 2.76 GeV respectively. We point out that none of the coupling constants
fixed in (\ref{meson-SU4-result}, \ref{baryon-SU4-result}) by a SU(4) constraint affect
the spectrum. As a consequence of the OZI rule only the t-channel exchange
of the light-vector mesons contribute. In the last two columns of Tab. \ref{tab:result1}
the spectrum is presented if we admit a small OZI violation in the coupling of the
light-vector mesons to the D-mesons. A finite coupling strength of the
$\phi$-meson to the D-mesons is permitted:
\begin{eqnarray}
{\mathcal L}^{\rm{SU(3)}}_{\rm{int}} &=&{\textstyle{i\over 2}}\,
\big( g_{\phi_\mu D \bar D}\,\phi^\mu
+g_{\omega_\mu D \bar D}\,\omega^\mu \big)\,
\big[  (\partial_\mu \bar D)\,D -\bar D \,(\partial_\mu D)
\big]
\nonumber\\
&+& {\textstyle{i\over 2}}\,g_{\rho_\mu D \bar D}
\,\rho^\mu \,\big[  (\partial_\mu \bar D)\,\vec \sigma \,D
-\bar D \,\vec \sigma \,(\partial_\mu D)
\big]\,,
\nonumber\\ \nonumber\\
&& g_{\phi_\mu D \bar D} = \sqrt{2}\,(h_{\bar 3\bar 3}^1+{\textstyle{1\over 2}}\,h^9_{\bar 3 \bar 3})\,,
\qquad g_{\omega_\mu D \bar D} = 2\,h_{\bar 3\bar 3}^1+{\textstyle{1\over 2}}\,h^9_{\bar 3 \bar 3}\,,
\nonumber\\
&& g_{\rho_\mu D \bar D} = {\textstyle{1\over 2}}\,h^9_{\bar 3 \bar 3}\,.
\end{eqnarray}
We introduce a small OZI violation by increasing the magnitude of $h_{\bar 3 \bar 3}^1 $ by about
19 $\%$ away from its SU(4) value. This keeps the value
$g_{\rho_\mu D \bar D} =g = 6.6$ but dials $g_{\phi_\mu D \bar D} \neq 0 $. With
$g_{\phi_\mu D \bar D} \simeq -1.74 $  the binding energy of the two bound states
is reduced by 93 MeV and 65 MeV as shown in Tab. \ref{tab:result1}.


\subsection{S-wave resonances with zero charm}

\begin{figure}[t]
\begin{center}
\includegraphics[clip=true,width=14.cm]{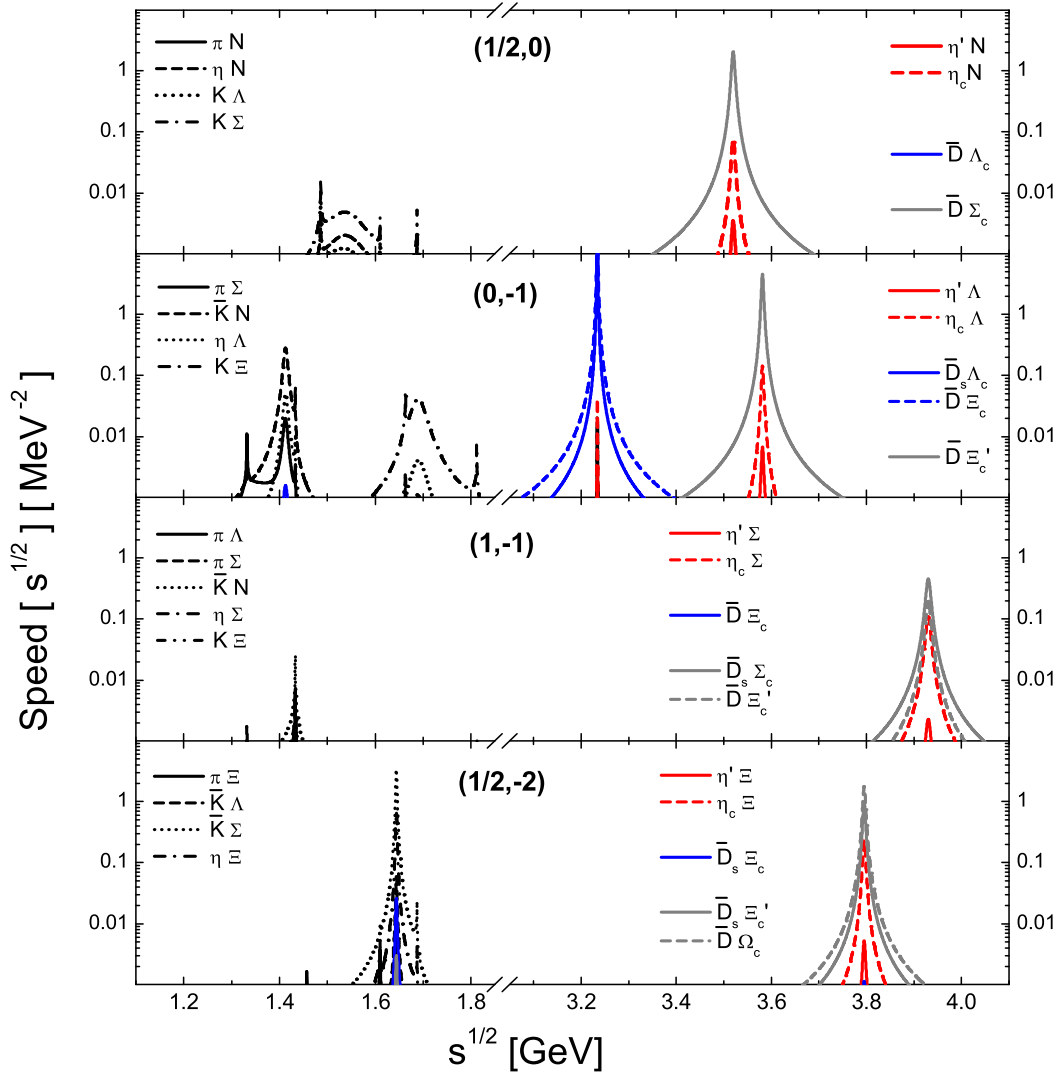}
\caption{It is shown the logarithm of the diagonal Speed$_{aa}(\sqrt{s}\,)$ for all
channels where resonances form with $C=0$. }
\label{fig:speed1}
\end{center}
\end{figure}

We turn to the resonances with $C=0$. The spectrum as shown in Fig. \ref{fig:speed1}
in terms of the speed introduced in (\ref{def-speed}), falls into two types of states.
Resonances with masses above 3 GeV couple strongly to mesons with non-zero charm content.
In the SU(3) limit those states form an octet and a singlet.
All other states have masses below 2 GeV.
In the SU(3) limit they group into two degenerate octets and one singlet.
The presence of the heavy channels does not affect that part of the spectrum at all.
This is reflected in coupling constants of those states to the heavy channels within
the typical range of $g \sim 0.1$ (see Tabs. \ref{tab:charm0a}-\ref{tab:charm0b}). We reproduce the
success of previous coupled-channel
computations \cite{Granada,Copenhagen}, which predicts the existence of the s-wave resonances
$N(1535), \Lambda(1405), \Lambda(1670),\Xi(1690)$ unambiguously with masses and branching ratios
quite compatible with empirical information. There are some quantitative differences.
This is the consequence of the t-channel vector meson exchange, which, only in the SU(3) limit
with degenerate vector meson masses, is equivalent to the Weinberg-Tomozawa interaction the
computation in \cite{Granada,Copenhagen} was based on.

Most spectacular are the resonances with hidden charm above 3 GeV.
The multiplet structure of
such states is readily understood. The mesons with $C=-1$  form a triplet which is scattered off
the $C=+1$ baryons forming a anti-triplet or sextet. We decompose the products into irreducible tensors
\begin{eqnarray}
3 \otimes \overline 3 = 1 \oplus 8 \,, \qquad
3 \otimes 6 = 8\oplus 10\,.
\label{3times6}
\end{eqnarray}
The matrix $\sum_{V\in [9]}\,C_V$ of (\ref{def-K}) is attractive in
the singlet for the triplet of baryons. Attraction in the octet sector is provided
by the sextet of baryons. The resulting octet of states mixes with the
$\eta'\,(N,\Lambda, \Sigma, \Xi)$ and
$\eta_c\,(N,\Lambda, \Sigma, \Xi)$ systems. A complicated mixing pattern arises.
All together the binding energies
of the crypto-exotic states are large. This is in part due to the large masses of
the coupled-channel states: the kinetic energy the attractive t-channel force has to
overcome is reduced. A second kinematical effect, which  further increases the binding energy, is
implied by the specific form of the t-channel exchange (\ref{VWT}). It provides the factor
$2\,\sqrt{s}-M-\bar M$. If evaluated at threshold it scales with the meson mass.

\begin{table}[t]
\rescale
\setlength{\tabcolsep}{1.2mm}
\setlength{\arraycolsep}{2.2mm}
\renewcommand{\arraystretch}{0.75}
\begin{center}
\begin{tabular}{|ll|c|c|c|c|c|}
\hline $C=0:$ &$ (\,I,\phantom{+}S)$  &
$\rm state$ &  $\begin{array}{c} M_R [\rm MeV]  \\ \Gamma_R \,[\rm MeV]  \end{array}$ &
$|g_R|$  & $\begin{array}{c} M_R [\rm MeV]  \\ \Gamma_R \,[\rm MeV]  \end{array}$ & $|g_R|$  \\
\hline
\hline
&$(\frac12,\phantom{+}0)$   &
$\begin{array}{l}  \pi\, N \\ \eta \,N \\ K \,\Lambda \\ K \,\Sigma \\
\eta' N \\ \eta_c N \\ \bar{D} \,\Lambda_c \\ \bar{D}\, \Sigma_c \end{array}$
& $\begin{array}{c} 1535 \\ 95 \end{array}$ & $\begin{array}{c} 0.3 \\ 2.1 \\ 1.7 \\ 3.3 \\ 0.0 \\ 0.0 \\ 0.2 \\ 0.2 \end{array}$
& $\begin{array}{c} 1536 \\ 94 \end{array}$ & $\begin{array}{c} 0.3 \\ 2.1 \\ 1.6 \\ 3.3 \\ 0.0 \\ 0.0 \\ 0.2 \\ 0.2 \end{array}$\\
\cline{3-7}
&$(\frac12,\phantom{+}0)$   &
$\begin{array}{l}  \pi\, N \\ \eta\, N \\ K \,\Lambda \\ K \,\Sigma \\
\eta' N \\ \eta_c N \\ \bar{D}\, \Lambda_c \\ \bar{D}\, \Sigma_c \end{array}$
& $\begin{array}{c} 3327 \\ 156 \end{array}$ & $\begin{array}{c} 0.1  \\ 0.1  \\ 0.1  \\ 0.1  \\ 1.4  \\ 0.7 \\ 0.5  \\ 5.7 \end{array}$
& $\begin{array}{c} 3520 \\ 7.3 \end{array}$ & $\begin{array}{c} 0.07 \\ 0.11 \\ 0.08 \\ 0.08 \\ 0.22 \\ 1.0 \\ 0.05 \\ 5.3 \end{array}$\\
\hline
&$(0,-1)$    &
$\begin{array}{l} \pi \,\Sigma \\ \bar{K}\, N \\ \eta\, \Lambda \\ K\, \Xi \\
\eta' \Lambda \\ \eta_c \Lambda \\ \bar{D}_s \Lambda_c \\ \bar{D}\, \Xi_c \\
\bar{D} \,\Xi_c'\end{array}$
& $\begin{array}{c} 1413 \\ 10 \end{array}$ & $\begin{array}{c} 0.7 \\ 2.7 \\ 1.1 \\ 0.1 \\ 0.0 \\ 0.0 \\ 0.2 \\ 0.0 \\ 0.0 \end{array}$
& $\begin{array}{c} 1413 \\ 10 \end{array}$ & $\begin{array}{c} 0.7 \\ 2.7 \\ 1.1 \\ 0.1 \\ 0.0 \\ 0.0 \\ 0.2 \\ 0.0 \\ 0.0 \end{array}$\\
\cline{3-7}
&$(0,-1)$    &
$\begin{array}{l} \pi\, \Sigma \\ \bar{K}\, N \\ \eta \,\Lambda \\ K\, \Xi \\ \eta' \Lambda \\
\eta_c \Lambda \\ \bar{D}_s \Lambda_c \\ \bar{D}\, \Xi_c \\ \bar{D}\, \Xi_c'\end{array}$
& $\begin{array}{c} 1689 \\ 35 \end{array}$ & $\begin{array}{c} 0.2 \\ 0.6 \\ 1.1 \\ 3.6 \\ 0.0 \\ 0.0 \\ 0.1 \\ 0.1 \\ 0.1 \end{array}$
& $\begin{array}{c} 1689 \\ 35 \end{array}$ & $\begin{array}{c} 0.2 \\ 0.6 \\ 1.1 \\ 3.6 \\ 0.0 \\ 0.0 \\ 0.1 \\ 0.1 \\ 0.1 \end{array}$\\
\cline{3-7}
&$(0,-1)$    &
$\begin{array}{l} \pi \,\Sigma \\ \bar{K}\, N \\ \eta\, \Lambda \\ K \,\Xi \\ \eta' \Lambda \\
\eta_c \Lambda \\ \bar{D}_s \Lambda_c \\ \bar{D}\, \Xi_c \\ \bar{D}\, \Xi_c'\end{array}$
& $\begin{array}{c} 3148 \\ 1.0  \end{array}$ & $\begin{array}{c} 0.04 \\ 0.03 \\ 0.03 \\ 0.04 \\ 0.08 \\ 0.08 \\ 3.2 \\ 5.0 \\ 0.1  \end{array}$
& $\begin{array}{c} 3234 \\ 0.57 \end{array}$ & $\begin{array}{c} 0.04 \\ 0.03 \\ 0.03 \\ 0.04 \\ 0.01 \\ 0.06 \\ 3.0 \\ 5.0 \\ 0.01 \end{array}$\\
\hline
\end{tabular}
\caption{
Spectrum of $J^P=\frac{1}{2}^-$ baryons with charm zero. The 3rd and 4th columns follow
with SU(4) symmetric 3-point vertices. In the
5th and 6th columns SU(4) breaking is introduced with  $h_{\bar 3 \bar 3}^1  \simeq -1.19 \,g$ and
$h^{\bar 3}_{ \bar 3 1 }=h^{\bar 3}_{1 \bar 3  } \simeq 0.71 \,g $. We use $g=6.6$.}
\label{tab:charm0a}
\end{center}
\end{table}

\begin{table}[t]
\rescale
\setlength{\tabcolsep}{1.2mm}
\setlength{\arraycolsep}{2.2mm}
\renewcommand{\arraystretch}{0.75}
\begin{center}
\begin{tabular}{|ll|c|c|c|c|c|}
\hline $C=0:$ &$ (\,I,\phantom{+}S)$  &
$\rm state$ &  $\begin{array}{c} M_R [\rm MeV]  \\ \Gamma_R \,[\rm MeV]  \end{array}$ &
$|g_R|$  & $\begin{array}{c} M_R [\rm MeV]  \\ \Gamma_R \,[\rm MeV]  \end{array}$ & $|g_R|$  \\
\hline
\hline
&$(0,-1)$    &
$\begin{array}{l} \pi\, \Sigma \\ \bar{K}\, N \\ \eta \,\Lambda \\ K \,\Xi \\
\eta' \Lambda \\ \eta_c \Lambda \\ \bar{D}_s \Lambda_c \\ \bar{D} \,\Xi_c \\ \bar{D}\, \Xi_c'\end{array}$
& $\begin{array}{c} 3432 \\ 161 \end{array}$ & $\begin{array}{c} 0.1  \\ 0.0  \\ 0.0  \\ 0.1  \\ 1.3  \\ 0.7  \\ 0.6  \\ 0.1  \\ 5.6 \end{array}$
& $\begin{array}{c} 3581 \\ 4.9 \end{array}$ & $\begin{array}{c} 0.06 \\ 0.01 \\ 0.03 \\ 0.07 \\ 0.20 \\ 0.93 \\ 0.05 \\ 0.02 \\ 5.3 \end{array}$\\
\hline
&$(1,-1)$    &
$\begin{array}{l} \pi \,\Lambda \\ \pi\, \Sigma \\ \bar{K}\, N \\ \eta\, \Sigma \\ K \,\Xi \\
\eta' \Sigma \\ \eta_c \Sigma \\ \bar{D}\, \Xi_c \\ \bar{D}_s \Sigma_c \\ \bar{D}\, \Xi_c' \end{array} $
& $\begin{array}{c} 3602 \\ 227 \end{array}$ & $\begin{array}{c} 0.1   \\ 0.1  \\ 0.2  \\ 0.1  \\ 0.1  \\ 1.5  \\ 1.2 \\ 0.6  \\ 4.6 \\ 2.9 \end{array}$
& $\begin{array}{c} 3930 \\ 11  \end{array}$ & $\begin{array}{c} 0.08  \\ 0.04 \\ 0.12 \\ 0.08 \\ 0.06 \\ 0.27 \\ 1.8 \\ 0.11 \\ 3.6 \\ 2.4 \end{array}$ \\
\hline
&$(\frac12,-2)$  &
$\begin{array}{l} \pi\, \Xi \\ \bar{K}\, \Lambda \\ \bar{K}\, \Sigma \\ \eta\, \Xi \\ \eta' \Xi \\
\eta_c \Xi \\ \bar{D}_s \Xi_c \\ \bar{D}_s \Xi_c' \\ \bar{D}\, \Omega_c \end{array}$
& $\begin{array}{c} 1644 \\ 3.0 \end{array}$ & $\begin{array}{c}  0.1 \\ 0.4 \\ 2.8 \\ 1.3 \\ 0.0 \\ 0.0 \\ 0.2 \\ 0.1 \\ 0.0 \end{array}$
& $\begin{array}{c} 1644 \\ 3.1 \end{array}$ & $\begin{array}{c}  0.1 \\ 0.4 \\ 2.8 \\ 1.3 \\ 0.0 \\ 0.0 \\ 0.2 \\ 0.1 \\ 0.0 \end{array}$ \\
\cline{3-7}
&$(\frac12,-2)$  &
$\begin{array}{l} \pi\, \Xi \\ \bar{K}\, \Lambda \\ \bar{K}\, \Sigma \\ \eta\, \Xi \\ \eta' \Xi \\
\eta_c \Xi \\ \bar{D}_s \Xi_c \\ \bar{D}_s \Xi_c' \\ \bar{D}\, \Omega_c \end{array}$
& $\begin{array}{c} 3624 \\ 204 \end{array}$ & $\begin{array}{c} 0.1  \\ 0.1  \\ 0.1  \\ 0.0  \\ 1.4  \\ 1.0 \\ 0.6  \\ 3.3 \\ 4.3 \end{array}$
& $\begin{array}{c} 3798 \\ 6.0 \end{array}$ & $\begin{array}{c} 0.08 \\ 0.04 \\ 0.04 \\ 0.01 \\ 0.22 \\ 1.2 \\ 0.10 \\ 2.9 \\ 4.0 \end{array}$ \\
\hline
\end{tabular}
\caption{Continuation of Tab. \ref{tab:charm0a}.}
\label{tab:charm0b}
\end{center}
\end{table}

The states shown in Fig. \ref{fig:speed1} are narrow as a result of the OZI rule. The mechanism
is analogous to the one explaining the long life time of the $J/\Psi$-meson. We should mention,
however, a caveat. It turns out that the width of the crypto-exotic states is quite sensitive to
the presence of channels involving the $\eta'$ meson. This is a natural result
since the $\eta'$ meson is closely related to the $U_A(1)$ anomaly giving it large gluonic
components. The latter work against the OZI rule. The relevant interaction terms are readily
identified
\begin{eqnarray}
&& {\mathcal L}_{\rm{int}}^{\rm{SU(3)}} = {\textstyle{ i\over \sqrt{6}}}\,
( 3\,h^{\bar 3}_{ \bar 3 1 }-h^{\bar 3}_{ \bar 3 9 })\,
\big[ \bar D\,D^\mu + \bar D_sD^\mu_s  \big](\partial_\mu\eta')+h.c.
\nonumber\\
&& \qquad \quad -\,{\textstyle{ i\over \sqrt{6}}}\,
(h^{\bar 3}_{9 \bar 3 }- 3\,h^{\bar 3}_{1 \bar 3  })\,
\big[ (\partial_\mu \bar D)\, D^\mu + (\partial_\mu\bar D_s)\,D^\mu_s  \big]\,\eta'+h.c. \,.
\label{etap-coupling}
\end{eqnarray}
We emphasize that switching off the t-channel exchange of charm or using the SU(4) estimate
for the latter, strongly bound crypto-exotic states are formed. In Tabs. \ref{tab:charm0a}-\ref{tab:charm0b}
the zero-charm spectrum insisting on the SU(4) estimates
(\ref{meson-SU4-result}, \ref{baryon-SU4-result}) is shown in the 3rd and 4th
column. The mass of the crypto-exotic nucleon resonance comes at 3.33 GeV in this case.
Its width of 160 MeV is completely dominated by the $\eta' N$ decay. The properties
of that state can be adjusted easily to be consistent with the empirical values claimed
in \cite{Karnaukhov:91}.
The $\eta'$ coupling strength to the open-charm mesons can be turned off by decreasing the
magnitude of $h^{\bar 3}_{ \bar 3 1 }$ and $h^{\bar 3}_{ 1\bar 3 }$ by 33.3 $\%$ away
from their SU(4) values. As a result the width of the resonance is down to about 1-2 MeV
\footnote{We mention that the width of the crypto-exotic states can be reduced also by decreasing
the value of the coupling constant $g_{86}^{\bar 3}$. However, in order to obtain a width for
crypto-exotic nucleon resonance of 7 MeV that coupling constant had to be reduced by a factor
4-5.}. It is stressed that the masses of the crypto-exotic states are not affected at all.
The latter are increased most efficiently by allowing a OZI violating $\phi_\mu \,D \bar D $
vertex. We adjust $h_{\bar 3 \bar 3}^1  \simeq -1.19 \,g$  and
$h^{\bar 3}_{ \bar 3 1 }=h^{\bar 3}_{1 \bar 3  } \simeq 0.71 \,g $ as to obtain the
nucleon resonance mass and width at $3.52$ GeV and $7$ MeV. For all other parameters the
SU(4) estimates are used. The result of this choice of parameters is
shown in Fig. \ref{fig:speed1} and in the last two rows of Tabs. \ref{tab:charm0a}-\ref{tab:charm0b}.
Further  crypto-exotic  states, members of the aforementioned octet,
are predicted at mass 3.58 GeV $(0,-1)$ and 3.93 GeV $(1,-1)$. The multiplet
is completed with a $(\frac{1}{2},-2)$ state at 3.80 GeV. The decay widths of
these states center around $\sim 7$ MeV. This  reflects the
dominance of their decays into channels involving the $\eta'$ meson. The coupling constants
to the various channels are included in Tabs. \ref{tab:charm0a}-\ref{tab:charm0b}. They confirm the interpretation
that the crypto-exotic states discussed above are a consequence of a strongly attractive force
between the charmed mesons and the baryon sextet.

We continue with a discussion of the crypto-exotic SU(3) singlet state, which is formed due
to strong attraction in the $(\bar D_s\Lambda_c), (\bar D\,\Xi_c )$ system. Its
nature is quite different as compared to the one of the octet states. This is because
its coupling to the $\eta' \Lambda$ channel is largely suppressed. Indeed its width
is independent on the magnitude of $h^{\bar 3}_{ \bar 3 1 }=h^{\bar 3}_{1 \bar 3  }$
as demonstrated in Tabs. \ref{tab:charm0a}-\ref{tab:charm0b}. We identify this state with a signal claimed in the
$K^-p$ reaction, where a narrow hyperon state with 3.17 GeV mass and width smaller than 20 MeV
was seen \cite{Amirzadeh:79}. Using values for the coupling constants as suggested by SU(4) the
state has a mass and width of 3.148 GeV and 1 MeV (see 3rd and 4th column of
Tabs. \ref{tab:charm0a}-\ref{tab:charm0b}).
Using our favored parameter set with $h_{\bar 3 \bar 3}^1  \simeq -1.19 \,g$  and
$h^{\bar 3}_{ \bar 3 1 }=h^{\bar 3}_{1 \bar 3  } \simeq 0.71 \,g $ the binding energy is
decreased by about 80 MeV. The width is slightly reduced.

It is instructive to compute the formation cross section for the
$N(3520)$ and $\Lambda(3170)$. Providing
the factor $2/3$ to (\ref{formation-cross-section}) as required for the $\pi^- p$ initial state,
the relevant coupling constant of Tabs. \ref{tab:charm0a}-\ref{tab:charm0b} implies
\begin{eqnarray}
\sigma_{\pi^- p \,\to N(3520)} = \frac{1\, {\rm MeV}}{\Gamma_{N(3520)} }\,600\,\mu {\rm barn}\,,
\label{production-cross-N}
\end{eqnarray}
for a given total decay width. This is to be compared with the total cross section
$\sigma^{\rm tot}_{\pi^-p}\simeq  25$ mbarn at $\sqrt{s}\simeq 3.5$ GeV \cite{PDG04}.
The formation cross section in a $K^-p$ reaction  of a the crypto-exotic singlet state is
\begin{eqnarray}
\sigma_{K^- p \,\to \Lambda(3170)} =
\frac{1\, {\rm MeV}}{\Gamma_{\Lambda(3170)} }\,36\,\mu {\rm barn}\,,
\label{production-cross-N}
\end{eqnarray}
for a given total decay width. This is to be compared with the total cross section
$\sigma^{\rm tot}_{\bar K^-p}\simeq  25$ mbarn at $\sqrt{s}\simeq 3.2$ GeV \cite{PDG04}.
The empirical width is reported to be smaller
than 20 MeV \cite{Amirzadeh:79}. Note that the initial $K^-p$ state requires an additional factor
$1/4$ in (\ref{formation-cross-section}).

\clearpage

\subsection{S-wave resonances with charm one}

\begin{figure}[b]
\begin{center}
\includegraphics[clip=true,width=14.cm]{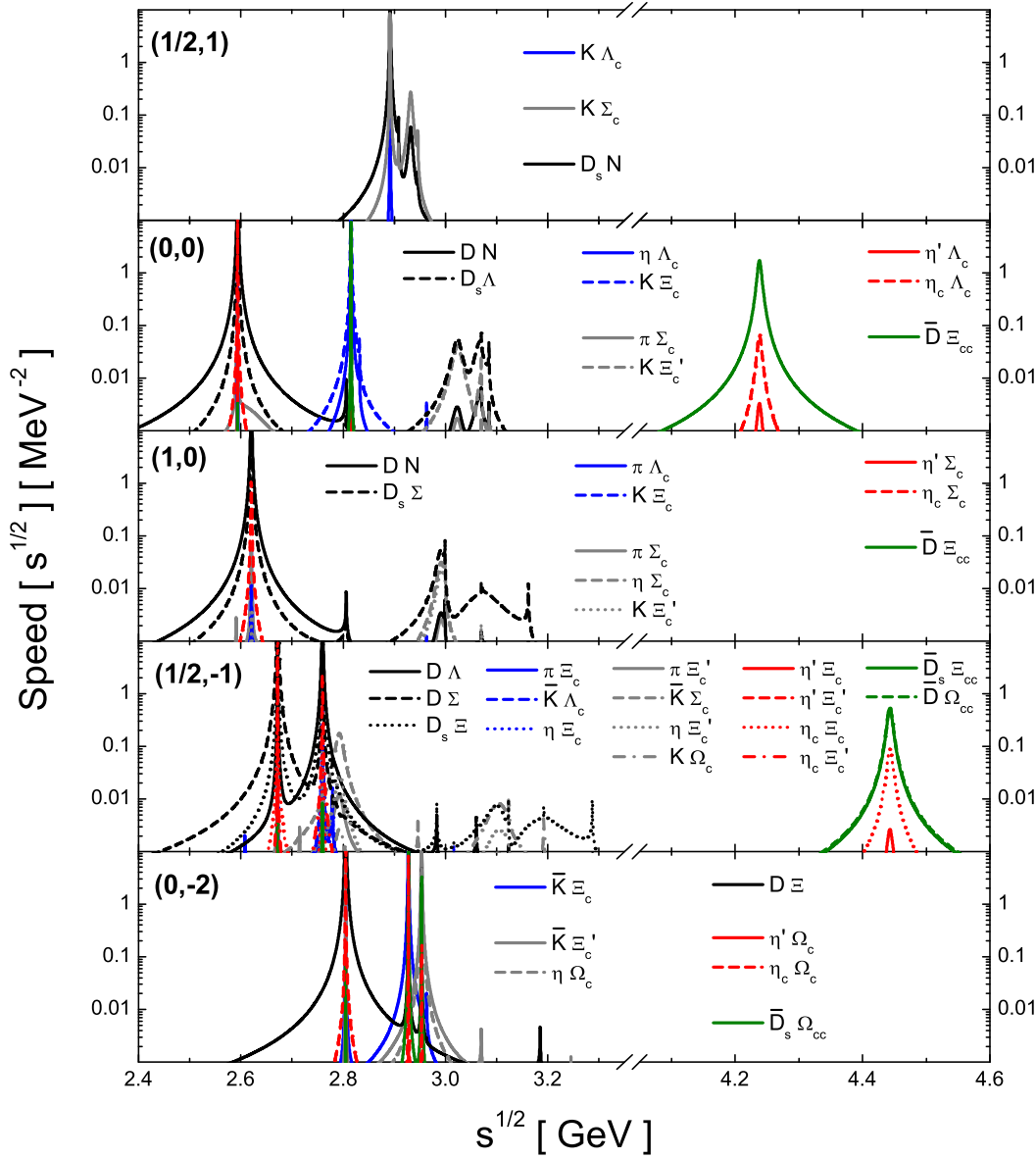}
\caption{It is shown the logarithm of the diagonal Speed$_{aa}(\sqrt{s}\,)$ for
channels where resonances form with $C=1$. }
\label{fig:speed2}
\end{center}
\end{figure}

Open charm systems are quite intriguing since the channels
which have either a charmed baryon or a charmed meson are comparatively close in mass.
Unfortunately, at present  there is very little empirical
information available on open-charm s-wave resonances. Only two states
$\Lambda_c(2593), \Xi_c(2790)$ \cite{PDG04} are discovered so far. We claim
that the $\Lambda_c(2880)$ observed by the CLEO collaboration \cite{Artuso}
can not be a s-wave resonance. This will be substantiated below.
In the speeds of Fig. \ref{fig:speed2} more than 15  well defined resonance states are
visible. Their properties are listed in Tabs. \ref{tab:charm1a}-\ref{tab:charm1d}. We collect the
resonance properties for the two parameter sets used before in the charm minus one and zero
sectors. The 3rd and 4th column show the implications of SU(4) symmetric 3-point vertices together
with the universal vector coupling constant $g =6.6$. The 5th and 6th columns
follow with the moderate SU(4) breaking relations $h_{\bar 3 \bar 3}^1  \simeq -1.19 \,g$  and
$h^{\bar 3}_{ \bar 3 1 }=h^{\bar 3}_{1 \bar 3  } \simeq 0.71 \,g $. The speeds shown in Fig.
\ref{fig:speed2} are based on the second set of parameters.

In previous coupled-channel computations the effect of
the light pseudo-scalar mesons as they scatter off charmed baryons was studied
\cite{LK04-charm,LK05}. We confirm the striking prediction of such
computations which suggest the existence of strongly bound $\bar 3, 6$ but also weakly
bound $\bar 3, 6,\overline{15}$ systems. These multiplets are formed by
scattering the octet of Goldstone bosons off the  baryon anti-triplet and sextet.
\begin{eqnarray}
&& 8 \otimes \bar 3= \bar 3 \oplus 6 \oplus \overline{15} \,,\qquad
8 \otimes 6= \bar 3 \oplus 6 \oplus \overline{15}\oplus 24\,.
\label{8times3}
\end{eqnarray}
For $8 \otimes \bar 3$ scattering chiral dynamics predicts attraction in anti-triplet, sextet
and repulsion in the $\overline {15}$-plet \cite{LK04-charm}. For $8 \otimes 6$ scattering
attraction is foreseen in anti-triplet, sextet, $\overline{15}$-plet with decreasing strength.
Further multiplets are generated by the scattering of the anti-triplet mesons of the octet
baryons. The decomposition is given already in (\ref{8times3}). In this case we find attraction in
the anti-triplet, sextet and the $\overline{15}$-plet. If we switch off the t-channel forces defined by the exchange
of heavy vector mesons the three types of resonances discussed above do not communicate with each
other. This is a direct consequence of the chiral SU(3) symmetry imposed in
(\ref{chiral-constraint-baryons}). It forbids the transition of a anti-triplet baryon into a
sextet baryon under the radiation of a light vector meson. Since the exchange of heavy vector
mesons is largely suppressed, the SU(4) assumption in (\ref{mesons-SU4}, \ref{baryons-SU4}) has
a very minor effect on the
resonance spectrum. The coupling constants (\ref{meson-SU4-result}, \ref{baryon-SU4-result})
estimate the small mixing of the three types of states.
For the readers'
convenience we recall the $(I,S)$ content of the various SU(3) multiplets:
\begin{eqnarray}
&& [\bar 3] \ni \left( \begin{array}{c} (0,0) \\ (\frac{1}{2},-1) \end{array}\right)\,, \qquad
[6] \ni \left(
\begin{array}{c}  (1,0) \\ (\frac{1}{2},-1) \\
(0,-2)\end{array}
\right) \,,\qquad
\nonumber\\
&& [\overline{15}] \ni
\left(
\begin{array}{c}
(\frac{1}{2},+1)\\
(0,0),(1,0)\\
(\frac{1}{2},-1),(\frac{3}{2},-1)\\
(1,-2)
\end{array}\right)\,,\qquad
 [24] \ni
\left(
 \begin{array}{c}
(\frac{3}{2},+1)\\
(1,0),(2,0)\\
(\frac{1}{2},-1),(\frac{3}{2},-1)\\
(0,-2),(1,-2)\\
(\frac{1}{2},-3)
\end{array}
\right) \,.
\label{recall-multiplets}
\end{eqnarray}

\begin{table}[t]
\rescale
\setlength{\tabcolsep}{1.2mm}
\setlength{\arraycolsep}{2.2mm}
\renewcommand{\arraystretch}{0.75}
\begin{center}
\begin{tabular}{|ll|c|c|c|c|c|}
\hline $C=1:$ &$ (\,I,\phantom{+}S)$  &
$\rm state$ &  $\begin{array}{c} M_R [\rm MeV]  \\ \Gamma_R \,[\rm MeV]  \end{array}$ &
$|g_R|$  & $\begin{array}{c} M_R [\rm MeV]  \\ \Gamma_R \,[\rm MeV]  \end{array}$ & $|g_R|$  \\
\hline
\hline
&$(\frac12,\phantom{+}1)$   &
$\begin{array}{l}  K\, \Lambda_c \\ D_s N \\ K \,\Sigma_c \end{array}$
& $\begin{array}{c} 2925 \\ 1.3 \end{array}$ & $\begin{array}{c}  0.02 \\ 0.4 \\ 2.0 \end{array}$
& $\begin{array}{c} 2932 \\ 6.9 \end{array}$ & $\begin{array}{c}  0.1  \\ 0.8 \\ 1.8 \end{array}$ \\
\cline{3-7}
&$(\frac12,\phantom{+}1)$   &
$\begin{array}{l}  K \,\Lambda_c \\ D_s N \\ K \,\Sigma_c
\end{array}$
&---&
& $\begin{array}{c} 2892 \\ 0.6 \end{array}$ & $\begin{array}{c}  0.1  \\ 2.9 \\ 0.9 \end{array}$ \\
\hline
&$(0,\phantom{+}0)$     &
$\begin{array}{l}  \pi\, \Sigma_c \\ D\, N \\ \eta \,\Lambda_c \\ K \,\Xi_c \\ K \,\Xi_c' \\
D_s \Lambda \\ \eta' \Lambda_c \\ \eta_c \Lambda_c \\ \bar{D}\, \Xi_{cc}   \end{array}$
& $\begin{array}{c} 2609 \\ 0.2  \end{array}$ & $\begin{array}{c}  0.14 \\ 6.3 \\ 0.04 \\ 0.01 \\ 0.03 \\ 2.7 \\ 0.11 \\ 0.62 \\ 0.0 \end{array}$
& $\begin{array}{c} 2593 \\ 0.05 \end{array}$ & $\begin{array}{c}  0.12 \\ 6.6 \\ 0.04 \\ 0.01 \\ 0.02 \\ 2.6 \\ 0.01 \\ 0.55 \\ 0.0 \end{array}$ \\
\cline{3-7}
&$(0,\phantom{+}0)$     &
$\begin{array}{l}  \pi \,\Sigma_c \\ D \,N \\ \eta \,\Lambda_c \\ K\, \Xi_c \\ K\, \Xi_c' \\
D_s \Lambda \\ \eta' \Lambda_c \\ \eta_c \Lambda_c \\ \bar{D}\, \Xi_{cc}   \end{array}$
& $\begin{array}{c} 2815 \\ 0.001  \end{array}$ & $\begin{array}{c}  0.0 \\ 0.01 \\ 1.3 \\ 2.4 \\ 0.0 \\ 0.11 \\ 0.0 \\ 0.02 \\ 0.15 \end{array}$
& $\begin{array}{c} 2815 \\ 0.0001 \end{array}$ & $\begin{array}{c}  0.0 \\ 0.0  \\ 1.3 \\ 2.4 \\ 0.0 \\ 0.15 \\ 0.0 \\ 0.02 \\ 0.16 \end{array}$ \\
\cline{3-7}
&$(0,\phantom{+}0)$     &
$\begin{array}{l}  \pi\, \Sigma_c \\ D\, N \\ \eta\, \Lambda_c \\ K \,\Xi_c \\ K \,\Xi_c' \\
D_s \Lambda \\ \eta' \Lambda_c \\ \eta_c \Lambda_c \\ \bar{D} \,\Xi_{cc}   \end{array}$
& $\begin{array}{c} 3036 \\ 17 \end{array}$ & $\begin{array}{c}  0.5 \\ 0.1 \\ 0.0 \\ 0.0 \\ 2.2 \\ 0.4 \\ 0.0 \\ 0.0 \\ 0.0 \end{array}$
& $\begin{array}{c} 3023 \\ 19 \end{array}$ & $\begin{array}{c}  0.4 \\ 0.5 \\ 0.1 \\ 0.1 \\ 1.9 \\ 2.2 \\ 0.0 \\ 0.1 \\ 0.0 \end{array}$ \\
\cline{3-7}
&$(0,\phantom{+}0)$     &
$\begin{array}{l}  \pi \,\Sigma_c \\ D \,N \\ \eta \,\Lambda_c \\ K\, \Xi_c \\ K \,\Xi_c' \\
D_s \Lambda \\ \eta' \Lambda_c \\ \eta_c \Lambda_c \\ \bar{D}\, \Xi_{cc}   \end{array}$
&---&
& $\begin{array}{c} 3068 \\ 22 \end{array}$ & $\begin{array}{c}  0.3 \\ 0.7 \\ 0.1 \\ 0.1 \\ 0.8 \\ 2.4 \\ 0.0 \\ 0.1 \\ 0.0 \end{array}$ \\
\cline{3-7}
&$(0,\phantom{+}0)$     &
$\begin{array}{l}  \pi\, \Sigma_c \\ D \,N \\ \eta \,\Lambda_c \\ K\, \Xi_c \\ K \Xi_c' \\
D_s \Lambda \\ \eta' \Lambda_c \\ \eta_c \Lambda_c \\ \bar{D} \,\Xi_{cc}   \end{array}$
& $\begin{array}{c} 4102 \\ 208 \end{array}$ & $\begin{array}{c}  0.1  \\ 0.1  \\ 0.0  \\ 0.1  \\ 0.0  \\ 0.1 \\ 1.3  \\ 0.7  \\ 4.9 \end{array}$
& $\begin{array}{c} 4238 \\ 7.3 \end{array}$ & $\begin{array}{c}  0.04 \\ 0.02 \\ 0.03 \\ 0.10 \\ 0.03 \\ 0.0 \\ 0.21 \\ 0.94 \\ 4.8 \end{array}$ \\
\hline
\end{tabular}
\caption{Spectrum of $J^P=\frac{1}{2}^-$ baryons with charm one. The 3rd and 4th columns follow
with SU(4) symmetric 3-point vertices. In the
5th and 6th columns SU(4) breaking is introduced with  $h_{\bar 3 \bar 3}^1  \simeq -1.19 \,g$ and
$h^{\bar 3}_{ \bar 3 1 }=h^{\bar 3}_{1 \bar 3  } \simeq 0.71 \,g $. We use $g=6.6$.}
\label{tab:charm1a}
\end{center}
\end{table}

\begin{table}[t]
\rescale
\setlength{\tabcolsep}{1.2mm}
\setlength{\arraycolsep}{2.2mm}
\renewcommand{\arraystretch}{0.75}
\begin{center}
\begin{tabular}{|ll|c|c|c|c|c|}
\hline $C=1:$ &$ (\,I,\phantom{+}S)$  &
$\rm state$ & $\begin{array}{c} M_R [\rm MeV]  \\ \Gamma_R \,[\rm MeV]  \end{array}$ &
$|g_R|$  & $\begin{array}{c} M_R [\rm MeV]  \\ \Gamma_R \,[\rm MeV]  \end{array}$ & $|g_R|$  \\
\hline
\hline
&$(1,\phantom{+}0)$     &
$\begin{array}{l}  \pi\, \Lambda_c \\ \pi \,\Sigma_c \\ D \,N \\ K\, \Xi_c \\ \eta \,\Sigma_c \\
K \,\Xi_c' \\ D_s \Sigma \\ \eta' \Sigma_c \\ \bar{D}\, \Xi_{cc} \\ \eta_c \Sigma_c  \end{array}$
& $\begin{array}{c} 2680 \\ 3.3 \end{array}$ & $\begin{array}{c} 0.2 \\ 0.2 \\ 4.9 \\ 0.1 \\ 0.0 \\ 0.1 \\ 3.6 \\ 0.1 \\ 0.0 \\ 0.8 \end{array}$
& $\begin{array}{c} 2620 \\ 1.4 \end{array}$ & $\begin{array}{c} 0.2 \\ 0.2 \\ 5.8 \\ 0.1 \\ 0.0 \\ 0.0 \\ 3.4 \\ 0.0 \\ 0.0 \\ 0.7 \end{array}$ \\
\cline{3-7}
&$(1,\phantom{+}0)$     &
$\begin{array}{l}  \pi \,\Lambda_c \\ \pi\, \Sigma_c \\ D \,N \\ K\, \Xi_c \\ \eta \,\Sigma_c \\
K \,\Xi_c' \\ D_s \Sigma \\ \eta' \Sigma_c \\ \bar{D}\, \Xi_{cc} \\ \eta_c \Sigma_c  \end{array}$
&---&
& $\begin{array}{c} 2992 \\ 18 \end{array}$ &
$\begin{array}{c} 0.0 \\ 0.5 \\ 0.5 \\ 0.0 \\ 1.6 \\ 1.4 \\ 2.1 \\ 0.0 \\ 0.1 \\ 0.1 \end{array}$ \\
\hline
&$(\frac12,-1)$  &
$\begin{array}{l}  \pi \,\Xi_c \\ \pi \,\Xi_c' \\ \bar{K}\, \Lambda_c \\ \bar{K}\, \Sigma_c \\
D \,\Lambda \\ \eta \,\Xi_c \\ D \,\Sigma \\ \eta\, \Xi_c' \\ K \,\Omega_c \\ D_s \Xi \\
\eta' \Xi_c \\ \eta' \Xi_c' \\ \eta_c \Xi_c \\ \bar D_s \Xi_{cc} \\ \bar{D}\, \Omega_{cc} \\
\eta_c \Xi_c'  \end{array}$
& $\begin{array}{c} 2691 \\ 0.09 \end{array}$ & $\begin{array}{c}  0.06 \\ 0.04 \\ 0.04 \\ 0.13 \\ 0.92 \\ 0.02 \\ 6.9 \\ 0.04 \\ 0.0 \\ 2.9 \\ 0.10 \\ 0.02 \\ 0.67 \\ 0.0 \\ 0.0 \\ 0.2 \end{array}$
& $\begin{array}{c} 2672 \\ 0.06 \end{array}$ & $\begin{array}{c}  0.05 \\ 0.03 \\ 0.03 \\ 0.10 \\ 0.93 \\ 0.02 \\ 7.1 \\ 0.03 \\ 0.0 \\ 2.8 \\ 0.01 \\ 0.0  \\ 0.60 \\ 0.0 \\ 0.01 \\ 0.19 \end{array}$ \\
\hline
\end{tabular}
\caption{Continuation of Tab. \ref{tab:charm1a}.}
\label{tab:charm1b}
\end{center}
\end{table}

\begin{table}[t]
\rescale
\setlength{\tabcolsep}{1.2mm}
\setlength{\arraycolsep}{2.2mm}
\renewcommand{\arraystretch}{0.75}
\begin{center}
\begin{tabular}{|ll|c|c|c|c|c|}
\hline $C=1:$ &$ (\,I,\phantom{+}S)$  &
$\rm state$ & $\begin{array}{c} M_R [\rm MeV]  \\ \Gamma_R \,[\rm MeV]  \end{array}$ &
$|g_R|$  & $\begin{array}{c} M_R [\rm MeV]  \\ \Gamma_R \,[\rm MeV]  \end{array}$ & $|g_R|$  \\
\hline
\hline
&$(\frac12,-1)$  &
$\begin{array}{l}  \pi\, \Xi_c \\ \pi\, \Xi_c' \\ \bar{K}\, \Lambda_c \\ \bar{K}\, \Sigma_c \\
D \,\Lambda \\ \eta \,\Xi_c \\ D \,\Sigma \\ \eta \,\Xi_c' \\ K \,\Omega_c \\ D_s \Xi \\
\eta' \Xi_c \\ \eta' \Xi_c' \\ \eta_c \Xi_c \\ \bar D_s \Xi_{cc} \\ \bar{D}\, \Omega_{cc} \\
\eta_c \Xi_c'  \end{array}$
& $\begin{array}{c} 2793 \\ 15  \end{array}$ & $\begin{array}{c}  0.0 \\ 0.9 \\ 0.1 \\ 3.4 \\ 0.0   \\ 0.1 \\ 0.1 \\ 1.3 \\ 0.4 \\ 0.0   \\ 0.0 \\ 0.0 \\ 0.0 \\ 0.1 \\ 0.0 \\ 0.0  \end{array}$
& $\begin{array}{c} 2793 \\ 16  \end{array}$ & $\begin{array}{c}  0.1 \\ 0.9 \\ 0.1 \\ 3.3 \\ 0.3 \\ 0.1 \\ 0.3 \\ 1.3 \\ 0.4 \\ 0.3 \\ 0.0 \\ 0.0 \\ 0.0 \\ 0.2 \\ 0.0 \\ 0.1 \end{array}$ \\
\cline{3-7}
&$(\frac12,-1)$  & $\begin{array}{l}  \pi\, \Xi_c \\ \pi\, \Xi_c' \\ \bar{K}\, \Lambda_c \\ \bar{K}\, \Sigma_c \\
D \,\Lambda \\ \eta \,\Xi_c \\ D \,\Sigma \\ \eta \,\Xi_c' \\ K \,\Omega_c \\ D_s \Xi \\
\eta' \Xi_c \\ \eta' \Xi_c' \\ \eta_c \Xi_c \\ \bar D_s \Xi_{cc} \\ \bar{D}\, \Omega_{cc} \\
\eta_c \Xi_c'  \end{array}$
& $\begin{array}{c} 2806 \\ 6.7 \end{array}$ & $\begin{array}{c} 0.3 \\ 0.1 \\ 0.4 \\ 0.7 \\ 4.8 \\ 0.3 \\ 1.8 \\ 0.2 \\ 0.1 \\ 3.7 \\ 0.1 \\ 0.1 \\ 0.3 \\ 0.0 \\ 0.0 \\ 0.8  \end{array}$
& $\begin{array}{c} 2759 \\ 0.9 \end{array}$ & $\begin{array}{c} 0.1 \\ 0.1 \\ 0.4 \\ 0.1 \\ 5.2 \\ 0.2 \\ 1.7 \\ 0.1 \\ 0.1 \\ 3.5 \\ 0.0 \\ 0.0 \\ 0.2 \\ 0.0 \\ 0.0 \\ 0.6  \end{array}$ \\
\cline{3-7}
&$(\frac12,-1)$  & $\begin{array}{l}  \pi\, \Xi_c \\ \pi\, \Xi_c' \\ \bar{K}\, \Lambda_c \\ \bar{K}\, \Sigma_c \\
D \,\Lambda \\ \eta \,\Xi_c \\ D \,\Sigma \\ \eta \,\Xi_c' \\ K \,\Omega_c \\ D_s \Xi \\
\eta' \Xi_c \\ \eta' \Xi_c' \\ \eta_c \Xi_c \\ \bar D_s \Xi_{cc} \\ \bar{D}\, \Omega_{cc} \\
\eta_c \Xi_c'  \end{array}$
& $\begin{array}{c} 3114 \\ 42 \end{array}$ & $\begin{array}{c} 0.0 \\ 0.7 \\ 0.0 \\ 0.4 \\ 0.2 \\ 0.0 \\ 0.1 \\ 1.1 \\ 2.1 \\ 0.4 \\ 0.0 \\ 0.0 \\ 0.0 \\ 0.0 \\ 0.1 \\ 0.0 \end{array}$
& $\begin{array}{c} 3104 \\ 43 \end{array}$ & $\begin{array}{c} 0.0 \\ 0.7 \\ 0.1 \\ 0.3 \\ 0.4 \\ 0.1 \\ 0.3 \\ 1.1 \\ 2.0 \\ 1.8 \\ 0.0 \\ 0.0 \\ 0.1 \\ 0.0 \\ 0.1 \\ 0.1 \end{array}$ \\
\hline
\end{tabular}
\caption{Continuation of Tab. \ref{tab:charm1b}.}
\label{tab:charm1c}
\end{center}
\end{table}

\begin{table}[t]
\rescale
\setlength{\tabcolsep}{1.2mm}
\setlength{\arraycolsep}{2.2mm}
\renewcommand{\arraystretch}{0.75}
\begin{center}
\begin{tabular}{|ll|c|c|c|c|c|}
\hline $C=1:$ &$ (\,I,\phantom{+}S)$  &
$\rm state$ & $\begin{array}{c} M_R [\rm MeV]  \\ \Gamma_R \,[\rm MeV]  \end{array}$ &
$|g_R|$  & $\begin{array}{c} M_R [\rm MeV]  \\ \Gamma_R \,[\rm MeV]  \end{array}$ & $|g_R|$  \\
\hline
\hline
&$(\frac12,-1)$  &
$\begin{array}{l}  \pi\, \Xi_c \\ \pi\, \Xi_c' \\ \bar{K}\, \Lambda_c \\ \bar{K}\, \Sigma_c \\
D\, \Lambda \\ \eta \,\Xi_c \\ D \,\Sigma \\ \eta \,\Xi_c' \\ K \,\Omega_c \\ D_s \Xi \\
\eta' \Xi_c \\ \eta' \Xi_c' \\ \eta_c \Xi_c \\ \bar D_s \Xi_{cc} \\ \bar{D}\, \Omega_{cc} \\
\eta_c \Xi_c'  \end{array}$
& $\begin{array}{c} 4274 \\ 241 \end{array}$ & $\begin{array}{c}  0.1 \\ 0.0 \\ 0.1 \\ 0.1 \\ 0.1 \\ 0.1 \\ 0.1 \\ 0.0 \\ 0.0 \\ 0.1 \\ 1.4 \\ 0.0 \\ 0.9 \\ 3.5 \\ 3.5 \\ 0.0 \end{array}$
& $\begin{array}{c} 4443 \\ 9   \end{array}$ & $\begin{array}{c}  0.1 \\ 0.0 \\ 0.1 \\ 0.0 \\ 0.0 \\ 0.1 \\ 0.0 \\ 0.0 \\ 0.0 \\ 0.0 \\ 0.2 \\ 0.0 \\ 1.3 \\ 3.2 \\ 3.3 \\ 0.0 \end{array}$\\
\hline \hline
&$(\frac32,-1)$  &
$\begin{array}{l} \pi \,\Xi_c \\ \pi \,\Xi_c' \\ \bar{K}\, \Sigma_c \\ D \,\Sigma \end{array}$
&---&
& $\begin{array}{c} 3052 \\ 15 \end{array}$ & $\begin{array}{c}  0.2 \\ 0.3 \\ 0.4 \\ 2.5 \end{array}$ \\
\hline \hline
&$(0,-2)$    &
$\begin{array}{l}  \bar{K}\, \Xi_c \\ \bar{K}\, \Xi_c' \\ D \,\Xi \\ \eta\, \Omega_c \\
\eta' \Omega_c \\ \bar D_s \Omega_{cc} \\ \eta_c \Omega_c \end{array}$
& $\begin{array}{c} 2839 \\ 0 \end{array}$ & $\begin{array}{c}  0.4 \\ 0.2 \\ 6.6 \\ 0.1 \\ 0.1 \\ 0.0 \\ 0.8 \end{array}$
& $\begin{array}{c} 2805 \\ 0 \end{array}$ & $\begin{array}{c}  0.3 \\ 0.1 \\ 6.9 \\ 0.1 \\ 0.0 \\ 0.0 \\ 0.7 \end{array}$ \\
\cline{3-7}
&$(0,-2)$    &
$\begin{array}{l}  \bar{K} \,\Xi_c \\ \bar{K} \,\Xi_c' \\ D\, \Xi \\ \eta \,\Omega_c \\
\eta' \Omega_c \\ \bar D_s \Omega_{cc} \\ \eta_c \Omega_c \end{array}$
& $\begin{array}{c} 2928 \\ 0 \end{array}$ & $\begin{array}{c}  2.3 \\ 0.1 \\ 0.7 \\ 0.1 \\ 0.0 \\ 0.4 \\ 0.1 \end{array}$
& $\begin{array}{c} 2927 \\ 0 \end{array}$ & $\begin{array}{c}  2.3 \\ 0.1 \\ 0.4 \\ 0.1 \\ 0.0 \\ 0.4 \\ 0.1 \end{array}$ \\
\cline{3-7}
&$(0,-2)$    &
$\begin{array}{l}  \bar{K}\, \Xi_c \\ \bar{K}\, \Xi_c' \\ D \,\Xi \\ \eta\, \Omega_c \\
\eta' \Omega_c \\ \bar D_s \Omega_{cc} \\ \eta_c \Omega_c \end{array}$
& $\begin{array}{c} 2953 \\ 0   \end{array}$ & $\begin{array}{c}  0.1 \\ 2.5 \\ 0.3 \\ 1.8 \\ 0.0 \\ 0.2 \\ 0.1 \end{array}$
& $\begin{array}{c} 2953 \\ 0   \end{array}$ & $\begin{array}{c}  0.0 \\ 2.5 \\ 0.2 \\ 1.8 \\ 0.0 \\ 0.2 \\ 0.1 \end{array}$ \\
\hline
\hline
&$(1,-2)$    &
$\begin{array}{l}  \pi\,\Omega_c \\ \bar{K}\, \Xi_c \\ \bar{K}\, \Xi_c' \\ D \,\Xi \end{array}$
&---&
& $\begin{array}{c} 3815 \\ 5 \end{array}$ & $\begin{array}{c}  0.2 \\ 0.2 \\ 0.2 \\ 1.6 \end{array}$ \\
\hline
\end{tabular}
\caption{Continuation of Tab. \ref{tab:charm1c}.}
\label{tab:charm1d}
\end{center}
\end{table}

All together from this discussion we expect the
formation of three strongly bound anti-triplet and sextet resonances and two weakly
bound $\overline{15}$-plets. It should be emphasized that coupled-channel dynamics tends to distort
the multiplet structure that arises in a SU(3) world. Not all members of a multiplet
will survive, in particular if there is weak attraction only.
The sextet resonances are most easily traced in the
$(0,-2)$ sector which is a unique signal of a sextet, given the fact that we
expect no 24 resonance. Indeed in this sector three bound states
with masses 2.80 GeV, 2.93 GeV and 2.95 GeV are displayed in
Fig. \ref{fig:speed2}. The coupling constants given in Tabs. \ref{tab:charm1a}-\ref{tab:charm1d} confirm
the above interpretation. There is a clear hierarchy of binding energies.
The states with large coupling constants to the anti-triplet mesons are bound most strongly.
The weakest binding is observed for states that couple strongly to the 6 baryons.
It is interesting to observe a distortion of that picture in the $(1,0)$ sector, in
which the sextet but also the $\overline{15}$-plet manifest themselves.
A strongly bound state at around 2.62 GeV couples dominantly to the $\bar 3$ mesons.
The second narrow state around 2.99 GeV is a member of the $\overline{15}$.
The chiral excitations of the anti-triplet and sextet baryons are quite broad in this
sector \cite{LK04-charm,LK05} and therefore not included in Tabs.
\ref{tab:charm1a}-\ref{tab:charm1d}. The $\overline{15}$-plet is clearly visible
also in the exotic $(\frac{1}{2},-1),(\frac{3}{2},-1),(1,-2)$ sectors with
narrow but weakly bound states. Depending on the parameter set we find one
or two states with $(\frac{1}{2},-1)$ whereas the $(\frac{3}{2},-1),(1,-2)$ sectors enjoy one
state only that couples strongly to the $(D\,\Sigma)$ and $(D_s\,\Xi)$ channels respectively.

The anti-triplet states are identified most easily in the $(0,0)$ sector.
The narrow state at 2.593 GeV couples strongly to the anti-triplet mesons. It has properties
amazingly consistent with the $\Lambda_c(2593)$ \cite{PDG04}. The empirical width is
$3.6^{+2.0}_{-1.3}$ MeV. This narrow state is almost degenerate in mass with a chiral excitation
of the triplet baryons \cite{LK04-charm,LK05}. The latter can be seen in Fig. \ref{fig:speed2} but
is not included in Tabs. \ref{tab:charm1a}-\ref{tab:charm1d}. It
decays dominantly into the $\pi\, \Sigma_c$ channel giving it a width of about 50 MeV.
A further narrow state at 2.815 GeV is the second chiral excitation of the
anti-triplet baryons \cite{LK04-charm,LK05} in this sector. Since it couples strongly
to the $\eta\,\Lambda_c(2285)$ channel, one should not associate this state with the
$\Lambda_c(2880)$ detected by the CLEO collaboration \cite{Artuso} via
its decay into the $\pi\, \Sigma_c(2453)$ channel.
The narrow total width of the observed state of smaller than 8 MeV \cite{Artuso} appears
inconsistent with a large coupling of that state to the open $\eta\,\Lambda_c$ channel.
The chiral excitation of the 6 baryon is quite broad in
this sector with mass around 2.65 GeV coupling strongly to the $\pi \Sigma_c$ channel.
Most spectacular is the $(\frac{1}{2},-1)$ sector in which we predict 4 narrow states below 4
GeV. The particle data group reports a state $\Xi_c(2790)$ with a decay width
smaller than 15 MeV. It is naturally identified with the chiral excitation of the sextet baryon of
mass 2.79 GeV and width 16 MeV seen in Fig. \ref{fig:speed2}. From Tabs. \ref{tab:charm1a}-\ref{tab:charm1d}
it follows that this state couples strongly to the $\bar K\,\Sigma_c$ and $\eta \,\Xi'_c$
channels.

Crypto-exotic states with $cc\bar c$ content are
formed by the scattering of the $3$-plet mesons with $C=-1$ off the triplet baryons with $C=2$:
\begin{eqnarray}
3 \otimes 3 = \bar 3 \oplus  6 \,,
\end{eqnarray}
where we predict strong attraction in the anti-triplet sector only.
The associated narrow states have masses ranging from 4.1 GeV to 4.4 GeV. Like in the case
of the crypto-exotic states in the zero-charm sectors these states decay preferably into
channels involving the $\eta'$. Depending on the parameter set the widths is large about
200-250 MeV or down to few MeV.

\subsection{S-wave resonances with charm two}

Baryon systems with $C=2$ are very poorly understood at present. There is a single published
isospin doublet state claimed by the SELEX collaboration at 3519 MeV \cite{SELEX} that
carries zero strangeness. There are hints that this state
can not be the ground state with $J^P=\frac{1}{2}^+ $ quantum numbers \cite{SELEX:Russ}.
This is why we postulated the somewhat ad-hoc mass of 3440 MeV for the ground state $\Xi_{cc}$.
Similarly the results of this section depend on our assumption for the $\Omega_{cc}$ mass
guessed at 3560 MeV.

\begin{figure}[t]
\begin{center}
\includegraphics[clip=true,width=14.cm]{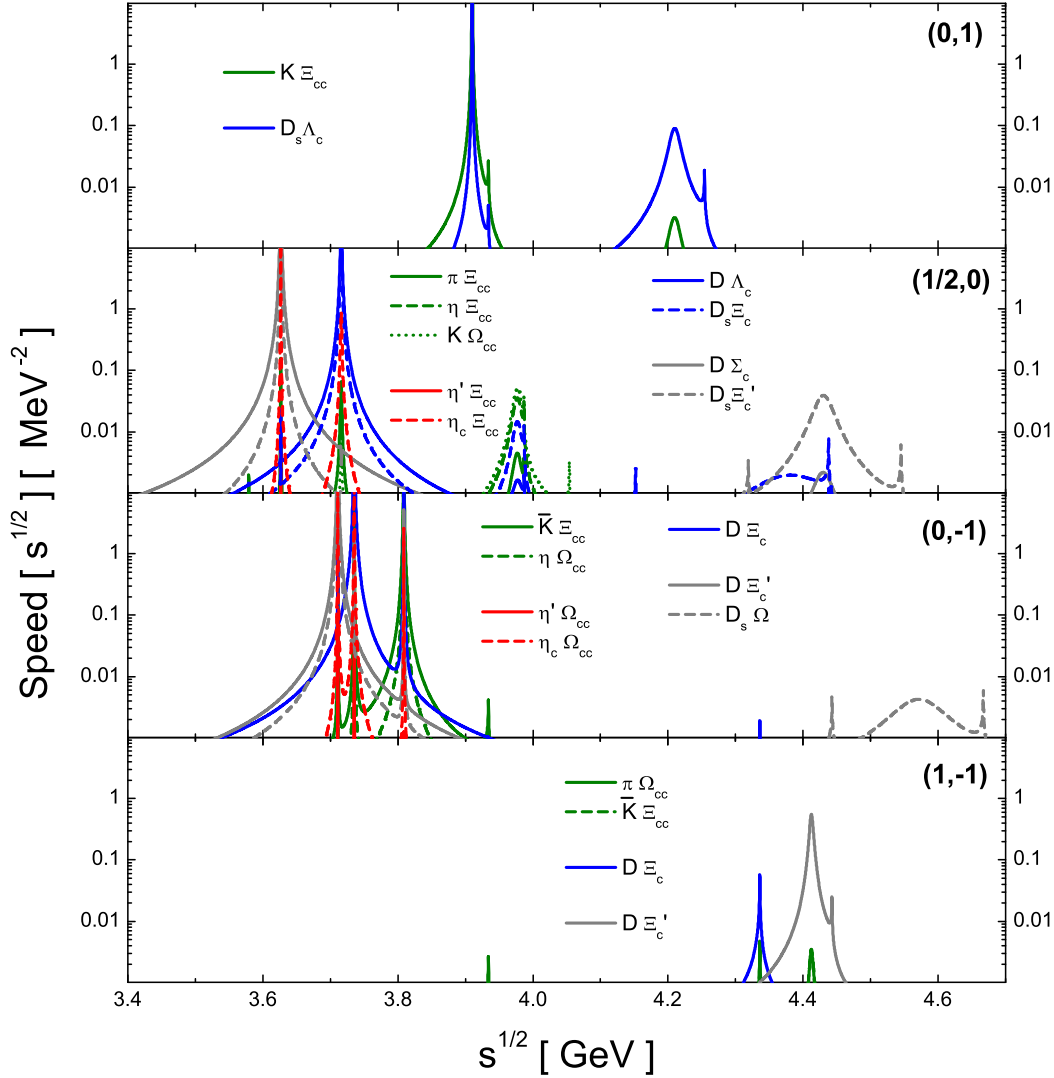}
\caption{It is shown the logarithm of the diagonal Speed$_{aa}(\sqrt{s}\,)$ for
channels where resonances form with $C=2$. }
\label{fig:speed3}
\end{center}
\end{figure}

There are three types of molecules formed in the
coupled-channel computations. The chiral excitations of the 3 baryons with $C=2$ form a
strongly bound triplet resonances and a less bound sextet resonance. This part
of the spectrum is analogous
to the one of the chiral excitations of open-charm mesons: attraction
is predicted in  the triplet and anti-sextet sectors \cite{KL04,HL04}:
\begin{eqnarray}
8 \otimes 3 =3 \oplus \bar 6 \oplus 15 \,, \qquad \bar 3 \otimes \bar 3 = 3\oplus \bar 6 \,,
\qquad \bar 3 \otimes 6 = 3 \oplus 15 \,.
\label{decomp-cc}
\end{eqnarray}
Further molecules are formed by the systems composed of open-charm mesons and open-charm baryons.
The multiplet decomposition for the anti-triplet and sextet baryons is given also in (\ref{decomp-cc}).
Strong attraction is predicted in both cases in the triplet, whereas the higher multiplets sextet and
$\overline{15}$-plet enjoy weak attraction only.
For the readers' convenience we recall the isospin strangeness content of the various multiplets:
\begin{eqnarray}
 && [3] \ni \left( \begin{array}{c} (\frac{1}{2},0) \\  (0,-1) \end{array}\right)\,, \quad \!
[\bar 6] \ni \left(
\begin{array}{c}  (0,+1) \\ (\frac{1}{2},0) \\
(1,-1)\end{array}
\right) \,, \quad \!
[15] \ni \left(
\begin{array}{c}  (1,+1) \\ (\frac{1}{2},0)\,,(\frac{3}{2},0) \\
(0,-1)\,,(1,-1)\\
(\frac{1}{2},-2)\end{array}
\right)\,.
\end{eqnarray}

\begin{table}[t]
\rescale
\setlength{\tabcolsep}{1.2mm}
\setlength{\arraycolsep}{2.2mm}
\renewcommand{\arraystretch}{0.75}
\begin{center}
\begin{tabular}{|ll|c|c|c|c|c|}
\hline $C=2:$ &$ (\,I,\phantom{+}S)$  &
$\rm state$ & $\begin{array}{c} M_R [\rm MeV]  \\ \Gamma_R \,[\rm MeV]  \end{array}$ &
$|g_R|$  & $\begin{array}{c} M_R [\rm MeV]  \\ \Gamma_R \,[\rm MeV]  \end{array}$ & $|g_R|$  \\
\hline
\hline
&$(0,\phantom{+}1)$     &
$\begin{array}{l}  K\, \Xi_{cc} \\ D_s \Lambda_c  \end{array}$
& $\begin{array}{c} 3912 \\ 0 \end{array}$ & $\begin{array}{c}  2.0 \\ 0.5 \end{array}$
& $\begin{array}{c} 3910 \\ 0 \end{array}$ & $\begin{array}{c}  2.0 \\ 0.8 \end{array}$ \\
\cline{3-7}
&$(0,\phantom{+}1)$     &
$\begin{array}{l}  K \,\Xi_{cc} \\ D_s \Lambda_c  \end{array}$
& --- &
& $\begin{array}{c} 4210 \\ 18 \end{array}$ & $\begin{array}{c}  0.5 \\ 2.7 \end{array}$ \\
\hline
&$(1,\phantom{+}1)$     &
$\begin{array}{l}  K \,\Xi_{cc} \\ D_s \Sigma_c  \end{array}$
&---&
& $\begin{array}{c} 4300 \\ 3.5 \end{array}$ & $\begin{array}{c}  0.2 \\ 3.4 \end{array}$ \\
\hline
&$(\frac12,\phantom{+}0)$   &
$\begin{array}{l}  \pi \,\Xi_{cc} \\ \eta\, \Xi_{cc} \\ K\, \Omega_{cc} \\ D \,\Lambda_c \\
D\, \Sigma_c \\ \eta' \Xi_{cc} \\ D_s \Xi_c \\ D_s \Xi_c' \\ \eta_c \Xi_{cc}  \end{array}$
& $\begin{array}{c} 3641 \\ 0.06 \end{array}$ & $\begin{array}{c}  0.05 \\ 0.02 \\ 0.02 \\ 0.02 \\ 6.2 \\ 0.07 \\ 0.01 \\ 2.6 \\ 0.44 \end{array}$
& $\begin{array}{c} 3626 \\ 0.04 \end{array}$ & $\begin{array}{c}  0.05 \\ 0.02 \\ 0.01 \\ 0.02 \\ 6.3 \\ 0.01 \\ 0.01 \\ 2.6 \\ 0.41 \end{array}$ \\
\cline{3-7}
&$(\frac12,\phantom{+}0)$   &
$\begin{array}{l}  \pi\, \Xi_{cc} \\ \eta\, \Xi_{cc} \\ K\, \Omega_{cc} \\ D\, \Lambda_c \\
D \,\Sigma_c \\ \eta' \Xi_{cc} \\ D_s \Xi_c \\ D_s \Xi_c' \\ \eta_c \Xi_{cc}  \end{array}$
& $\begin{array}{c} 3759 \\ 1.9 \end{array}$ & $\begin{array}{c}  0.2 \\ 0.0 \\ 0.1 \\ 4.6 \\ 0.1 \\ 0.2 \\ 3.4 \\ 0.0 \\ 1.0 \end{array}$
& $\begin{array}{c} 3716 \\ 1.8 \end{array}$ & $\begin{array}{c}  0.2 \\ 0.0 \\ 0.1 \\ 5.0 \\ 0.1 \\ 0.0 \\ 3.3 \\ 0.0 \\ 0.8 \end{array}$\\
\cline{3-7}
&$(\frac12,\phantom{+}0)$   &
$\begin{array}{l}  \pi\, \Xi_{cc} \\ \eta \,\Xi_{cc} \\ K\, \Omega_{cc} \\ D\, \Lambda_c \\
D\, \Sigma_c \\ \eta' \Xi_{cc} \\ D_s \Xi_c \\ D_s \Xi_c' \\ \eta_c \Xi_{cc}  \end{array}$
& $\begin{array}{c} 3979 \\ 14 \end{array}$ & $\begin{array}{c}  0.5 \\ 1.3 \\ 1.5 \\ 0.1 \\ 0.0 \\ 0.0 \\ 0.5 \\ 0.1 \\ 0.0 \end{array}$
& $\begin{array}{c} 3977 \\ 14 \end{array}$ & $\begin{array}{c}  0.5 \\ 1.3 \\ 1.6 \\ 0.2 \\ 0.0 \\ 0.0 \\ 0.8 \\ 0.1 \\ 0.0 \end{array}$ \\
\cline{3-7}
&$(\frac12,\phantom{+}0)$   &
$\begin{array}{l}  \pi\, \Xi_{cc} \\ \eta\, \Xi_{cc} \\ K\, \Omega_{cc} \\ D \,\Lambda_c \\
D\, \Sigma_c \\ \eta' \Xi_{cc} \\ D_s \Xi_c \\ D_s \Xi_c' \\ \eta_c \Xi_{cc}  \end{array}$
&---&
& $\begin{array}{c} 4430 \\ 32 \end{array}$ & $\begin{array}{c}  0.0 \\ 0.2 \\ 0.2 \\ 0.1 \\ 0.8 \\ 0.0 \\ 0.0\\ 3.2 \\ 0.1  \end{array}$ \\
\hline
\end{tabular}
\caption{Spectrum of $J^P=\frac{1}{2}^-$ baryons with charm two. The 3rd and 4th columns follow
with SU(4) symmetric 3-point vertices. In the
5th and 6th columns SU(4) breaking is introduced with  $h_{\bar 3 \bar 3}^1  \simeq -1.19 \,g$ and
$h^{\bar 3}_{ \bar 3 1 }=h^{\bar 3}_{1 \bar 3  } \simeq 0.71 \,g $. We use $g=6.6$.}
\label{tab:charm2a}
\end{center}
\end{table}

\begin{table}[t]
\rescale
\setlength{\tabcolsep}{1.2mm}
\setlength{\arraycolsep}{2.2mm}
\renewcommand{\arraystretch}{0.75}
\begin{center}
\begin{tabular}{|ll|c|c|c|c|c|}
\hline $C=2:$ &$ (\,I,\phantom{+}S)$  &
$\rm state$ & $\begin{array}{c} M_R [\rm MeV]  \\ \Gamma_R \,[\rm MeV]  \end{array}$ &
$|g_R|$  & $\begin{array}{c} M_R [\rm MeV]  \\ \Gamma_R \,[\rm MeV]  \end{array}$ & $|g_R|$  \\
\hline
\hline
&$(\frac32,\phantom{+}0)$   &
$\begin{array}{l} \pi\, \Xi_{cc} \\ D\, \Sigma_c \end{array} $
&---&
& $\begin{array}{c} 4251 \\ 4.6 \end{array}$ & $\begin{array}{c}  0.2 \\ 2.9 \end{array}$ \\
\hline
&$(0,-1)$    &
$\begin{array}{l}   \bar K\, \Xi_{cc} \\ \eta \,\Omega_{cc} \\
D \,\Xi_c \\ D \,\Xi_c' \\ \eta' \Omega_{cc} \\ D_s \Omega_c \\ \eta_c \Omega_{cc} \end{array}$
& $\begin{array}{c} 3726 \\ 0 \end{array}$ & $\begin{array}{c}  0.1 \\ 0.0 \\ 0.1 \\ 5.5 \\ 0.1 \\ 4.0 \\ 0.5  \end{array}$
& $\begin{array}{c} 3711 \\ 0 \end{array}$ & $\begin{array}{c}  0.1 \\ 0.0 \\ 0.1 \\ 5.6 \\ 0.0 \\ 3.7 \\ 0.5  \end{array}$ \\
\cline{3-7}
&$(0,-1)$    &
$\begin{array}{l}   \bar K\, \Xi_{cc} \\ \eta\, \Omega_{cc} \\ D \,\Xi_c \\ D\, \Xi_c' \\
\eta' \Omega_{cc} \\ D_s \Omega_c \\ \eta_c \Omega_{cc} \end{array}$
& $\begin{array}{c} 3761 \\ 0 \end{array}$ & $\begin{array}{c}  0.6 \\ 0.3 \\ 5.9 \\ 0.0 \\ 0.1 \\ 0.0 \\ 0.9 \end{array}$
& $\begin{array}{c} 3735 \\ 0 \end{array}$ & $\begin{array}{c}  0.3 \\ 0.2 \\ 6.2 \\ 0.1 \\ 0.0 \\ 0.0 \\ 0.8 \end{array}$ \\
\cline{3-7}
&$(0,-1)$    &
$\begin{array}{l}  \bar K \,\Xi_{cc} \\ \eta\, \Omega_{cc} \\ D \,\Xi_c \\ D\, \Xi_c' \\
\eta' \Omega_{cc} \\ D_s \Omega_c \\ \eta_c \Omega_{cc} \end{array}$
& $\begin{array}{c} 3810 \\ 0 \end{array}$ & $\begin{array}{c}  2.6 \\ 1.2  \\ 1.1 \\ 0.2 \\ 0.0 \\ 0.2 \\ 0.2  \end{array}$
& $\begin{array}{c} 3809 \\ 0 \end{array}$ & $\begin{array}{c}  2.7 \\ 1.2  \\ 0.7 \\ 0.1 \\ 0.0 \\ 0.2 \\ 0.1  \end{array}$ \\
\cline{3-7}
&$(0,-1)$    &
$\begin{array}{l}  \bar K \,\Xi_{cc} \\ \eta\, \Omega_{cc} \\ D \,\Xi_c \\ D\, \Xi_c' \\
\eta' \Omega_{cc} \\ D_s \Omega_c \\ \eta_c \Omega_{cc} \end{array}$
&---&
& $\begin{array}{c} 4571 \\ 90 \end{array}$ & $\begin{array}{c}  0.2 \\ 0.3 \\ 0.1 \\ 1.2 \\ 0.0 \\ 2.9 \\ 0.2 \end{array}$ \\
\hline
&$(1,-1)$    &
$\begin{array}{l} \pi\, \Omega_{cc} \\ \bar K \,\Xi_{cc} \\ D\,\Xi_c \\ D \,\Xi_c'  \end{array} $
&---&
& $\begin{array}{c} 4412 \\ 6 \end{array}$ & $\begin{array}{c}  0.2 \\ 0.2 \\ 0.0 \\ 2.3 \end{array}$ \\
\hline
&$(\frac12,-2)$  &
$\begin{array}{l} \bar K \,\Omega_{cc} \\ D \,\Omega_{c}  \end{array} $
&---&
& $\begin{array}{c} 4562 \\ 4 \end{array}$ & $\begin{array}{c}  0.2 \\ 1.3 \end{array}$ \\
\hline
\end{tabular}
\caption{Continuation of Tab. \ref{tab:charm2a}.}
\label{tab:charm2b}
\end{center}
\end{table}

All together we expect 6 states with
$(\frac{1}{2},0)$ quantum numbers.
Given the postulated values for the masses of the $3$ baryons, the lowest
$(\frac{1}{2},0)$ state is predicted at mass around 3.63-3.64 GeV depending on the
parameter set. That state couples preferably to the $(D \,\Sigma_c)$ and $(D_s\Xi_c')$ states, but
very weakly to the $(D\,\Lambda_c)$ channel.
Therefore it should not be identified with the SELEX $\Xi_{cc}^+$ state at 3519 MeV, which was
observed by its decay into $\Lambda_c^+\,K^-\,\pi^+$ \cite{SELEX}. A further resonance of the
$(D\,\Lambda_c),(D_s\Xi_c)$ system is formed at 3.72-3.76 GeV. Its mass is much to high as to
associate it with the SELEX state. If the SELEX state had $\frac{1}{2}^-$ quantum numbers it
poses certainly a puzzle to us. The first chiral excitation with a width
of about 200 MeV is predicted at mass 3.65 GeV. It decays into the open $\pi \,\Xi_{cc}$ channel.
Due to its broad width that resonance is not listed in Tabs. \ref{tab:charm2a}-\ref{tab:charm2b}.
The narrow chiral excitation at around 3.98 GeV,
which couples strongly to the $(\eta\, \Xi_{cc}), (K\, \Omega_{cc})$ channels, reflects the
attraction predicted in the anti-sextet. Two further $(\frac{1}{2},0)$ states may arise from
the weak attraction of the open-charm mesons and open-charm baryons in the anti-sextet and
$15$-plet. The formation of those states depends on the details of the parameter choice. For SU(4)
symmetric 3-point vertices no clear signals are seen in that anti-sextet and 15-plet.
However, with the SU(4) breaking
pattern used before a broad anti-sextet state at 4.38 GeV, that decays into the open $(D\,\Lambda_c)$
channel, is formed. The 15-plet manifests itself with a state at 4.43 GeV of width about 30 MeV.

The hierarchy observed for the binding energies of triplet states with $(\frac{1}{2},0)$ quantum numbers
is confirmed by the $(0,-1)$ states belonging to the 3. We predict three bound states at
(3.71, 3.74, 3.81) GeV for the parameter set with SU(4) breaking. A further state, necessarily a
member of the 15-plet is predicted at 4.57 GeV with a width of 90 MeV.
We turn to the $(0,+1)$ and $(1,-1)$ sectors. The first probes uniquely the 6. Indeed we
find two states with $(0,+1)$ at 3.91 GeV and 4.21 GeV. Again the chiral excitation is bound
stronger than the state coupling formed by the interaction of the $D_s$ with the
$\Lambda_c$. In the $(1,-1)$ sector we expect three states.
A narrow signal is seen at 4.41 GeV only, which is a member of the 15-plet.
The two states belonging to
the two 6 multiplets are masked by coupled channel effects. No clear signals are predicted.
It remains to discuss the three exotic sectors $(1,+1),(\frac{3}{2},0),(\frac{1}{2},-2)$, all
probing exclusively the 15-plet.  As described above, the attraction in the 15-plet is
sufficiently strong
to form a resonance multiplet only, if a SU(4) breaking pattern is allowed. As shown in the
4th and 5th columns of Tabs. \ref{tab:charm2a}-\ref{tab:charm2b}, weakly bound states are
generated in the before mentioned sectors.

\subsection{S-wave resonances with charm three}

We close the result section with a presentation of baryon states with $C=3$. They are formed
by scattering the triplet baryons with $C=2$ of the anti-triplet mesons with $C=1$.
According to the decomposition $3 \otimes \bar 3=1 \oplus 8 $ only a SU(3) singlet or
octet may arise. As demonstrated
in Tab. \ref{tab:charm3} we predict attraction in the singlet only with a bound state of
mass 4.31-4.33 GeV depending on the parameter set.

\begin{table}[t]
\rescale
\setlength{\tabcolsep}{1.2mm}
\setlength{\arraycolsep}{2.2mm}
\renewcommand{\arraystretch}{0.75}
\begin{center}
\begin{tabular}{|ll|c|c|c|c|c|}
\hline $C=3:$ &$ (\,I,\phantom{+}S)$  &
$\rm state$ & $\begin{array}{c} M_R [\rm MeV]  \\ \Gamma_R \,[\rm MeV]  \end{array}$ &
$|g_R|$  & $\begin{array}{c} M_R [\rm MeV]  \\ \Gamma_R \,[\rm MeV]  \end{array}$ & $|g_R|$  \\
\hline
\hline
&$(0,\phantom{+}0)$     & $\begin{array}{l}  D \,\Xi_{cc} \\ D_s \Omega_{cc}  \end{array}$
& $\begin{array}{c} 4325 \\ 0 \end{array}$ & $\begin{array}{c}  5.5 \\ 2.8 \end{array}$
& $\begin{array}{c} 4308 \\ 0 \end{array}$ & $\begin{array}{c}  5.6 \\ 2.8 \end{array}$\\
\hline
\end{tabular}
\caption{Spectrum of $J^P=\frac{1}{2}^-$ baryons with charm three. The 3rd and 4th columns follow
with SU(4) symmetric 3-point vertices. In the
5th and 6th columns SU(4) breaking is introduced with  $h_{\bar 3 \bar 3}^1  \simeq -1.19 \,g$ and
$h^{\bar 3}_{ \bar 3 1 }=h^{\bar 3}_{1 \bar 3  } \simeq 0.71 \,g $. We use $g=6.6$.}
\label{tab:charm3}
\end{center}
\end{table}

\clearpage

\section{Summary}

We have performed a coupled-channel study of s-wave baryon resonances with charm $-1,0,1,2,3$.
A rich spectrum is predicted in terms of a t-channel force defined by the exchange of
light vector mesons. All relevant coupling constants are obtained from chiral and large-$N_c$
properties of QCD. Less relevant vertices related to the t-channel forces induced by the
exchange of charmed vector mesons  were estimated by applying SU(4) symmetry.
We pointed out that the decay process $D_+(2010) \to \pi_+ \,D_0(1865)$ and the
$\rho_+(770) \to \pi_0\,\pi_+$ can be described by a SU(4) symmetric vertex, where only moderate
SU(4) breaking effects are required. As an amusing byproduct it was demonstrated that the KSFR
relation \cite{KSFRa,KSFRb,Djukanovic} can be viewed as consequence of
a 3-point vector-pseudoscalar-meson vertex that is SU(4) symmetric. The results of this
work should be taken cautiously since it remains to study the effect of additional
terms in the interaction kernel.

Most spectacular is the prediction of narrow crypto-exotic baryons with charm zero forming below
4 GeV. Such states contain a $c \bar c$ pair. Their widths parameters are small due to the OZI rule,
like it is the case for the $J/\Psi$ meson. We predict an octet of crypto-exotic states which
decay dominantly into channels involving an $\eta'$ meson. An even stronger bound crypto-exotic
SU(3) singlet state is predicted to have a decay width of about 1 MeV only. We recover the masses
and widths of a crypto-exotic nucleon and hyperon
resonance suggested in high statistic bubble chamber experiments \cite{Amirzadeh:79,Karnaukhov:91}.
We confirm the expectation of Lipkin \cite{Lipkin:87} that penta-quark type states exists with
charm minus one. Binding is predicted only in systems with strangeness minus one and minus two.
In the charm one sector more than ten so far unobserved narrow states are predicted. Further
narrow s-wave states are foreseen in the charm two and three sectors.

Our predictions can be tested experimentally in part by existing collaborations
like SELEX or BELLE. We urge the QCD lattice community to perform unquenched simulations
in order to verify or disprove the existence of the predicted states. Such studies are of great
importance since they will shed more light on how confinement is realized in  nature. The
central question - what are the most relevant degrees of freedom responsible for the formation of
resonances in QCD - can be studied best in systems involving light and heavy quarks
simultaneously. The spectrum of open-charm baryons could be a topic of interest for the FAIR
project at GSI.

{\bfseries{Acknowledgments}}

M.F.M.L. acknowledges useful discussions with
J. Engelfried, E.E. Kolomeitsev and J. Russ.

\section{Appendix A}

\clearpage

\begin{table}
\rescale
\setlength{\tabcolsep}{1.2mm}
\setlength{\arraycolsep}{3.0mm}
\renewcommand{\arraystretch}{1.0}
\begin{center}
\begin{tabular}{|l|l|l|c|}
\hline
\makebox[1cm][l]{$(I,\phantom{x}S)$} & \makebox[1cm][l]{$(a,b)$} &
\makebox[1.cm][l]{$V$} & \makebox[9.8cm][l]{$C_{V,ab}^{(I,S,-1)}\qquad \qquad \quad $ charm minus one} \\
\hline
\hline
$(0,+0)$    &$(1,1)$    &$\rho$     &$\frac32\,g_{88}^{9-}\,h_{\bar3\bar3}^9 + \frac32\,g_{88}^{9+}\,h_{\bar3\bar3}^9$ \\
        &       &$\omega$   &$4\,g_{88}^1\,h_{\bar3\bar3}^1 + 2\,g_{88}^{9-}\,h_{\bar3\bar3}^1 + 2\,g_{88}^{9+}\,h_{\bar3\bar3}^1 + g_{88}^1\,h_{\bar3\bar3}^9 + \frac12\,g_{88}^{9-}\,h_{\bar3\bar3}^9 + \frac12\,g_{88}^{9+}\,h_{\bar3\bar3}^9$ \\
        &       &$\phi$     &$2\,g_{88}^1\,h_{\bar3\bar3}^1 - 2\,g_{88}^{9-}\,h_{\bar3\bar3}^1 + 2\,g_{88}^{9+}\,h_{\bar3\bar3}^1 + g_{88}^1\,h_{\bar3\bar3}^9 - g_{88}^{9-}\,h_{\bar3\bar3}^9 + g_{88}^{9+}\,h_{\bar3\bar3}^9$ \\
        &       &$J/\Psi$   &$g_{88}^0\,h_{\bar3\bar3}^0$ \\
\hline
\hline
$(1,+0)$    &$(1,1)$    &$\rho$     &$-\frac12\,g_{88}^{9-}\,h_{\bar3\bar3}^9 - \frac12\, g_{88}^{9+}\,h_{\bar3\bar3}^9$\\
        &       &$\omega$   &$4\,g_{88}^1\,h_{\bar3\bar3}^1 + 2\,g_{88}^{9-}\,h_{\bar3\bar3}^1 + 2\,g_{88}^{9+}\,h_{\bar3\bar3}^1 + g_{88}^1\,h_{\bar3\bar3}^9 + \frac12\,g_{88}^{9-}\,h_{\bar3\bar3}^9 + \frac12\,g_{88}^{9+}\,h_{\bar3\bar3}^9$\\
        &       &$\phi$     &$2\,g_{88}^1\,h_{\bar3\bar3}^1 - 2\,g_{88}^{9-}\,h_{\bar3\bar3}^1 + 2\,g_{88}^{9+}\,h_{\bar3\bar3}^1 + g_{88}^1\,h_{\bar3\bar3}^9 - g_{88}^{9-}\,h_{\bar3\bar3}^9 + g_{88}^{9+}\,h_{\bar3\bar3}^9$\\
\hline
\hline
\end{tabular}
\caption{The coupled-channel structure of the t-channel exchange in (\ref{def-K}).}
\label{tab:appendix:1}
\end{center}
\end{table}

\begin{table}
\rescale
\setlength{\tabcolsep}{1.2mm}
\setlength{\arraycolsep}{3.0mm}
\renewcommand{\arraystretch}{1.0}
\begin{center}
\begin{tabular}{|l|l|l|c|}
\hline
\makebox[1cm][l]{$(I,\phantom{x}S)$} & \makebox[1cm][l]{$(a,b)$} & \makebox[1.cm][l]{$V$} &
\makebox[9.8cm][l]{$C_{V,ab}^{(I,S,0)}\qquad \qquad \qquad \quad \; \;\,$ charm zero} \\
\hline
\hline
$(0,+1)$&$(1,1)$&$\rho$&$3\,g_{88}^{9-}\,h_{99}^9 + 3\,g_{88}^{9+}\,h_{99}^9 $\\
&&$\omega$&$-2\,g_{88}^1\,h_{99}^9 - g_{88}^{9-}\,h_{99}^9 - g_{88}^{9+}\,h_{99}^9$\\
&&$\phi$&$2\,g_{88}^1\,h_{99}^9 - 2\,g_{88}^{9-}\,h_{99}^9 + 2\,g_{88}^{9+}\,h_{99}^9$\\
\hline
\hline
$(1,+1)$&$(1,1)$&$\rho$&$-g_{88}^{9-}\,h_{99}^9 - g_{88}^{9+}\,h_{99}^9 $\\
&&$\omega$&$-2\,g_{88}^1\,h_{99}^9 - g_{88}^{9-}\,h_{99}^9 - g_{88}^{9+}\,h_{99}^9$\\
&&$\phi$&$2\,g_{88}^1\,h_{99}^9 - 2\,g_{88}^{9-}\,h_{99}^9 + 2\,g_{88}^{9+}\,h_{99}^9$\\
\hline
\hline
$(\frac12,+0)$&$(1,1)$&$\rho$&$4\,g_{88}^{9-}\,h_{99}^9 + 4\,g_{88}^{9+}\,h_{99}^9 $\\
\hline
\end{tabular}
\caption{Continuation of Tab. \ref{tab:appendix:1}.}
\label{tab:appendix:2}
\end{center}
\end{table}

\begin{table}
\rescale
\setlength{\tabcolsep}{1.2mm}
\setlength{\arraycolsep}{3.0mm}
\renewcommand{\arraystretch}{1.0}
\begin{center}
\begin{tabular}{|l|l|l|c|}
\hline
\makebox[1cm][l]{$(I,\phantom{x}S)$} & \makebox[1cm][l]{$(a,b)$} & \makebox[1.cm][l]{$V$} & \makebox[9.8cm][l]{$C_{V,ab}^{(I,S,0)}\qquad \qquad \qquad \quad \; \;\,$ charm zero} \\
\hline
\hline
$(0,-1)$&$(1,1)$&$\rho$&$8\,g_{88}^{9-}\,h_{99}^9 $\\
\cline{2-3}
&$(1,2)$&$K^*$&$\sqrt{6}\,g_{88}^{9-}\,h_{99}^9 - \sqrt{6}\,g_{88}^{9+}\,h_{99}^9$\\
\cline{2-3}
&$(1,4)$&$K^*$&$-\sqrt{6}\,g_{88}^{9-}\,h_{99}^9 - \sqrt{6}\,g_{88}^{9+}\,h_{99}^9$\\
\cline{2-3}
&$(1,8)$&$D^*$&$-\sqrt{\frac34}\,g_{8\bar3}^{\bar3}\,h_{\bar39}^{\bar3} - \sqrt{\frac34}\,g_{8\bar3}^{\bar3}\,h_{9\bar3}^{\bar3}$\\
\cline{2-3}
&$(1,9)$&$D^*$&$-\sqrt{\frac{3}{32}}\,g_{86}^{\bar3}\,h_{\bar39}^{\bar3} - \sqrt{\frac{3}{32}}\,g_{86}^{\bar3}\,h_{9\bar3}^{\bar3}$\\
\cline{2-3}
&$(2,2)$&$\rho$&$3\,g_{88}^{9-}\,h_{99}^9 + 3\,g_{88}^{9+}\,h_{99}^9 $\\
&&$\omega$&$2\,g_{88}^1\,h_{99}^9 + g_{88}^{9-}\,h_{99}^9 + g_{88}^{9+}\,h_{99}^9$\\
\hline
\end{tabular}
\caption{Continuation of Tab. \ref{tab:appendix:2}.}
\label{tab:appendix:3}
\end{center}
\end{table}

\begin{table}
\rescale
\setlength{\tabcolsep}{1.2mm}
\setlength{\arraycolsep}{3.0mm}
\renewcommand{\arraystretch}{1.0}
\begin{center}
\begin{tabular}{|l|l|l|c|}
\hline
\makebox[1cm][l]{$(I,\phantom{x}S)$} & \makebox[1cm][l]{$(a,b)$} & \makebox[1.cm][l]{$V$} & \makebox[9.8cm][l]{$C_{V,ab}^{(I,S,0)}\qquad \qquad \qquad \quad \; \;\,$ charm zero} \\
\hline
\hline
$(1,-1)$&$(2,3)$&$K^*$&$2\,g_{88}^{9-}\,h_{99}^9 - 2\,g_{88}^{9+}\,h_{99}^9$\\
\cline{2-3}
&$(2,5)$&$K^*$&$-2\,g_{88}^{9-}\,h_{99}^9 - 2\,g_{88}^{9+}\,h_{99}^9$\\
\cline{2-3}
&$(2,8)$&$D^*$&$-\frac14\,g_{8\bar3}^{\bar3}\,h_{\bar39}^{\bar3} - \frac14\,g_{8\bar3}^{\bar3}\,h_{9\bar3}^{\bar3}$\\
\cline{2-3}
&$(2,10)$&$D^*$&$-\frac14\,g_{86}^{\bar3}\,h_{\bar39}^{\bar3} - \frac14\,g_{86}^{\bar3}\,h_{9\bar3}^{\bar3}$\\
\cline{2-3}
&$(3,3)$&$\rho$&$-g_{88}^{9-}\,h_{99}^9 - g_{88}^{9+}\,h_{99}^9 $\\
&&$\omega$&$2\,g_{88}^1\,h_{99}^9 + g_{88}^{9-}\,h_{99}^9 + g_{88}^{9+}\,h_{99}^9$\\
&&$\phi$&$-2\,g_{88}^1\,h_{99}^9 + 2\,g_{88}^{9-}\,h_{99}^9 - 2\,g_{88}^{9+}\,h_{99}^9$\\
\hline
\hline
\end{tabular}
\caption{Continuation of Tab. \ref{tab:appendix:3}.}
\label{tab:appendix:4}
\end{center}
\end{table}

\begin{table}
\rescale
\setlength{\tabcolsep}{1.2mm}
\setlength{\arraycolsep}{3.0mm}
\renewcommand{\arraystretch}{1.0}
\begin{center}
\begin{tabular}{|l|l|l|c|}
\hline
\makebox[1cm][l]{$(I,\phantom{x}S)$} & \makebox[1cm][l]{$(a,b)$} & \makebox[1.cm][l]{$V$} & \makebox[9.8cm][l]{$C_{V,ab}^{(I,S,0)}\qquad \qquad \qquad \quad \; \;\,$ charm zero} \\
\hline
\hline
$(\frac12,-2)$
&$(2,3)$&$\rho$&$-2\,g_{88}^{9+}\,h_{99}^9 $\\
\cline{2-3}
&$(2,4)$&$K^*$&$-3\,g_{88}^{9-}\,h_{99}^9 + g_{88}^{9+}\,h_{99}^9$\\
\cline{2-3}
&$(2,7)$&$D^*$&$\frac1{4\sqrt6}\,g_{8\bar3}^{\bar3}\,h_{\bar39}^{\bar3} + \frac1{4\sqrt6}\,g_{8\bar3}^{\bar3}\,h_{9\bar3}^{\bar3}$\\
\cline{2-3}
&$(2,8)$&$D^*$&$-\sqrt{\frac{3}{32}}\,g_{86}^{\bar3}\,h_{\bar39}^{\bar3} - \sqrt{\frac{3}{32}}\,g_{86}^{\bar3}\,h_{9\bar3}^{\bar3}$\\
\cline{2-3}
&$(3,3)$&$\rho$&$4\,g_{88}^{9-}\,h_{99}^9 $\\
&&$\omega$&$2\,g_{88}^1\,h_{99}^9 + 2\,g_{88}^{9+}\,h_{99}^9$\\
&&$\phi$&$-2\,g_{88}^1\,h_{99}^9$\\
\hline
\hline
\end{tabular}
\caption{Continuation of Tab. \ref{tab:appendix:4}.}
\label{tab:appendix:5}
\end{center}
\end{table}

\begin{table}
\rescale
\setlength{\tabcolsep}{1.2mm}
\setlength{\arraycolsep}{3.0mm}
\renewcommand{\arraystretch}{1.0}
\begin{center}
\begin{tabular}{|l|l|l|c|}
\hline
\makebox[1cm][l]{$(I,\phantom{x}S)$} & \makebox[1cm][l]{$(a,b)$} & \makebox[1.cm][l]{$V$} & \makebox[9.8cm][l]{$C_{V,ab}^{(I,S,1)}\qquad \qquad \qquad \quad \; \;\,$ charm one} \\
\hline
\hline
$(\frac12,+1)$&$(1,1)$&$\omega$&$-2\,g_{\bar3\bar3}^1\,h_{99}^9 - g_{\bar3\bar3}^9\,h_{99}^9$\\
&&$\phi$&$2\,g_{\bar3\bar3}^1\,h_{99}^9$\\
\cline{2-3}
&$(1,2)$&$D^*$&$\frac14\,g_{8\bar3}^{\bar3}\,h_{\bar39}^{\bar3} + \frac14\,g_{8\bar3}^{\bar3}\,h_{9\bar3}^{\bar3}$\\
\cline{2-3}
&$(1,3)$&$\rho$&$\sqrt3\, g_{\bar36}^9\,h_{99}^9$\\
\cline{2-3}
&$(2,2)$&$\omega$&$-4\,g_{88}^1\,h_{\bar3\bar3}^1 - 2\,g_{88}^{9-}\,h_{\bar3\bar3}^1 - 2\,g_{88}^{9+}\,h_{\bar3\bar3}^1 - 2\,g_{88}^1\,h_{\bar3\bar3}^9 - g_{88}^{9-}\,h_{\bar3\bar3}^9 - g_{88}^{9+}\,h_{\bar3\bar3}^9$\\
&&$\phi$&$-2\,g_{88}^1\,h_{\bar3\bar3}^1 + 2\,g_{88}^{9-}\,h_{\bar3\bar3}^1 - 2\,g_{88}^{9+}\,h_{\bar3\bar3}^1$\\
&&$J/\Psi$&$-g_{88}^0\,h_{\bar3\bar3}^0$\\
\hline
\end{tabular}
\caption{Continuation of Tab. \ref{tab:appendix:5}.}
\label{tab:appendix:6}
\end{center}
\end{table}

\begin{table}
\rescale
\setlength{\tabcolsep}{1.2mm}
\setlength{\arraycolsep}{3.0mm}
\renewcommand{\arraystretch}{1.0}
\begin{center}
\begin{tabular}{|l|l|l|c|}
\hline
\makebox[1cm][l]{$(I,\phantom{x}S)$} & \makebox[1cm][l]{$(a,b)$} & \makebox[1.cm][l]{$V$} & \makebox[9.8cm][l]{$C_{V,ab}^{(I,S,1)}\qquad \qquad \qquad \quad \; \;\,$ charm one} \\
\hline
\hline
$(0,+0)$&$(6,7)$&$D^*_s$&$-\sqrt{\frac{1}{2}}\,g_{8\bar3}^{\bar3}\,h_{1\bar3}^{\bar3} - \sqrt{\frac{1}{2}}\,g_{8\bar3}^{\bar3}\,h_{\bar31}^{\bar3} + \sqrt{\frac{1}{18}}\,g_{8\bar3}^{\bar3}\,h_{\bar39}^{\bar3} + \sqrt{\frac{1}{18}}\,g_{8\bar3}^{\bar3}\,h_{9\bar3}^{\bar3}$\\
\cline{2-3}
&$(6,8)$&$D^*_s$&$\sqrt{\frac{1}{12}}\,g_{8\bar3}^{\bar3}\,h_{0\bar3}^{\bar3} + \sqrt{\frac{1}{12}}\,g_{8\bar3}^{\bar3}\,h_{\bar30}^{\bar3}$\\
\cline{2-3}
&$(7,9)$&$D^*$&$-\sqrt{\frac{3}{4}}\,g_{\bar33}^{\bar3}\,h_{1\bar3}^{\bar3} - \sqrt{\frac{3}{4}}\,g_{\bar33}^{\bar3}\,h_{\bar31}^{\bar3} + \sqrt{\frac{1}{12}}\,g_{\bar33}^{\bar3}\,h_{\bar39}^{\bar3} + \sqrt{\frac{1}{12}}\,g_{\bar33}^{\bar3}\,h_{9\bar3}^{\bar3}$\\
\cline{2-3}
&$(8,9)$&$D^*$&$\sqrt{\frac{1}{8}}\,g_{\bar33}^{\bar3}\,h_{0\bar3}^{\bar3} + \sqrt{\frac{1}{8}}\,g_{\bar33}^{\bar3}\,h_{\bar30}^{\bar3}$\\
\cline{2-3}
&$(9,9)$&$\rho$&$-\frac34\,g_{33}^9\,h_{\bar3\bar3}^9 $\\
&&$\omega$&$4\,g_{33}^1\,h_{\bar3\bar3}^1 + g_{33}^9\,h_{\bar3\bar3}^1 + g_{33}^1\,h_{\bar3\bar3}^9 + \frac14\,g_{33}^9\,h_{\bar3\bar3}^9$\\
&&$\phi$&$2\,g_{33}^1\,h_{\bar3\bar3}^1 + g_{33}^9\,h_{\bar3\bar3}^1 + g_{33}^1\,h_{\bar3\bar3}^9 + \frac12\,g_{33}^9\,h_{\bar3\bar3}^9$\\
\hline
\end{tabular}
\caption{Continuation of Tab. \ref{tab:appendix:6}.}
\label{tab:appendix:7}
\end{center}
\end{table}

\begin{table}
\rescale
\setlength{\tabcolsep}{1.2mm}
\setlength{\arraycolsep}{3.0mm}
\renewcommand{\arraystretch}{1.0}
\begin{center}
\begin{tabular}{|l|l|l|c|}
\hline
\makebox[1cm][l]{$(I,\phantom{x}S)$} & \makebox[1cm][l]{$(a,b)$} & \makebox[1.cm][l]{$V$} & \makebox[9.8cm][l]{$C_{V,ab}^{(I,S,1)}\qquad \qquad \qquad \quad \; \;\,$ charm one} \\
\hline
\hline
$(1,+0)$
&$(7,10)$&$D^*_s$&$\frac12\,g_{86}^{\bar3}\,h_{0\bar3}^{\bar3} + \frac12\,g_{86}^{\bar3}\,h_{\bar30}^{\bar3}$\\
\cline{2-3}
&$(8,9)$&$D^*$&$-\sqrt3\,g_{63}^{\bar3}\,h_{1\bar3}^{\bar3} - \sqrt3\,g_{63}^{\bar3}\,h_{\bar31}^{\bar3} + \sqrt{\frac{1}{3}}\,g_{63}^{\bar3}\,h_{\bar39}^{\bar3} + \sqrt{\frac{1}{3}}\,g_{63}^{\bar3}\,h_{9\bar3}^{\bar3}$\\
\cline{2-3}
&$(9,9)$&$\rho$&$\frac14\,g_{33}^9\,h_{\bar3\bar3}^9 $\\
&&$\omega$&$4\,g_{33}^1\,h_{\bar3\bar3}^1 + g_{33}^9\,h_{\bar3\bar3}^1 + g_{33}^1\,h_{\bar3\bar3}^9 + \frac14\,g_{33}^9\,h_{\bar3\bar3}^9$\\
&&$\phi$&$2\,g_{33}^1\,h_{\bar3\bar3}^1 + g_{33}^9\,h_{\bar3\bar3}^1 + g_{33}^1\,h_{\bar3\bar3}^9 + \frac12\,g_{33}^9\,h_{\bar3\bar3}^9$\\
&&$J/\Psi$&$g_{33}^0\,h_{\bar3\bar3}^0$\\
\cline{2-3}
&$(9,10)$&$D^*$&$\sqrt{\frac{1}{2}}\,g_{63}^{\bar3}\,h_{0\bar3}^{\bar3} + \sqrt{\frac{1}{2}}\,g_{63}^{\bar3}\,h_{\bar30}^{\bar3}$\\
\hline
\end{tabular}
\caption{Continuation of Tab. \ref{tab:appendix:7}.}
\label{tab:appendix:8}
\end{center}
\end{table}

\begin{table}
\rescale
\setlength{\tabcolsep}{1.2mm}
\setlength{\arraycolsep}{3.0mm}
\renewcommand{\arraystretch}{1.0}
\begin{center}
\begin{tabular}{|l|l|l|c|}
\hline
\makebox[1cm][l]{$(I,\phantom{x}S)$} & \makebox[1cm][l]{$(a,b)$} & \makebox[1.cm][l]{$V$} & \makebox[9.8cm][l]{$C_{V,ab}^{(I,S,1)}\qquad \qquad \qquad \quad \; \;\,$ charm one} \\
\hline
\hline
$(\frac12,-1)$
&$(5,16)$&$D^*$&$-\sqrt{\frac{3}{16}}\,g_{86}^{\bar3}\,h_{0\bar3}^{\bar3} - \sqrt{\frac{3}{16}}\,g_{86}^{\bar3}\,h_{\bar30}^{\bar3}$\\
\cline{2-3}
&$(6,7)$&$D^*$&$\frac18\,g_{8\bar3}^{\bar3}\,h_{\bar39}^{\bar3} + \frac18\,g_{8\bar3}^{\bar3}\,h_{9\bar3}^{\bar3}$\\
\cline{2-3}
&$(6,9)$&$K^*$&$\sqrt3\,g_{\bar36}^9\,h_{99}^9 $\\
\cline{2-3}
&$(6,10)$&$D^*_s$&$\sqrt{\frac{1}{24}}\,g_{8\bar3}^{\bar3}\,h_{\bar39}^{\bar3} + \sqrt{\frac{1}{24}}\,g_{8\bar3}^{\bar3}\,h_{9\bar3}^{\bar3}$\\
\cline{2-3}
&$(6,14)$&$D^*_s$&$-\sqrt{\frac{1}{48}}\,g_{\bar33}^{\bar3}\,h_{\bar39}^{\bar3} - \sqrt{\frac{1}{48}}\,g_{\bar33}^{\bar3}\,h_{9\bar3}^{\bar3}$\\
\cline{2-3}
&$(6,15)$&$D^*$&$-\frac1{8\sqrt3}\,g_{\bar33}^{\bar3}\,h_{\bar39}^{\bar3} - \frac1{8\sqrt3}\,g_{\bar33}^{\bar3}\,h_{9\bar3}^{\bar3}$\\
\cline{2-3}
&$(7,7)$&$\rho$&$2\,g_{88}^{9-}\,h_{\bar3\bar3}^9 $\\
\hline
\hline
\end{tabular}
\caption{Continuation of Tab. \ref{tab:appendix:8}.}
\label{tab:appendix:9}
\end{center}
\end{table}

\begin{table}
\rescale
\setlength{\tabcolsep}{1.2mm}
\setlength{\arraycolsep}{3.0mm}
\renewcommand{\arraystretch}{1.0}
\begin{center}
\begin{tabular}{|l|l|l|c|}
\hline
\makebox[1cm][l]{$(I,\phantom{x}S)$} & \makebox[1cm][l]{$(a,b)$} & \makebox[1.cm][l]{$V$} & \makebox[9.8cm][l]{$C_{V,ab}^{(I,S,1)}\qquad \qquad \qquad \quad \; \;\,$ charm one} \\
\hline
\hline
$(\frac32,-1)$
&$(1,1)$&$\rho$&$-g_{\bar3\bar3}^9\,h_{99}^9 $\\
\cline{2-3}
&$(1,2)$&$\rho$&$-g_{\bar36}^9\,h_{99}^9 $\\
\cline{2-3}
&$(1,3)$&$K^*$&$-\sqrt2\,g_{\bar36}^9\,h_{99}^9 $\\
\cline{2-3}
&$(1,4)$&$D^*$&$-\frac14\,g_{8\bar3}^{\bar3}\,h_{\bar39}^{\bar3} - \frac14\,g_{8\bar3}^{\bar3}\,h_{9\bar3}^{\bar3}$\\
\cline{2-3}
&$(2,2)$&$\rho$&$-g_{66}^9\,h_{99}^9$\\
\cline{2-3}
&$(2,3)$&$K^*$&$\sqrt{2}\,g_{66}^9\,h_{99}^9$\\
\cline{2-3}
&$(2,4)$&$D^*$&$-\frac14\,g_{86}^{\bar3}\,h_{\bar39}^{\bar3} - \frac14\,g_{86}^{\bar3}\,h_{9\bar3}^{\bar3}$\\
\hline
\hline
\end{tabular}
\caption{Continuation of Tab. \ref{tab:appendix:9}.}
\label{tab:appendix:10}
\end{center}
\end{table}

\begin{table}
\rescale
\setlength{\tabcolsep}{1.2mm}
\setlength{\arraycolsep}{3.0mm}
\renewcommand{\arraystretch}{1.0}
\begin{center}
\begin{tabular}{|l|l|l|c|}
\hline
\makebox[1cm][l]{$(I,\phantom{x}S)$} & \makebox[1cm][l]{$(a,b)$} & \makebox[1.cm][l]{$V$} & \makebox[9.8cm][l]{$C_{V,ab}^{(I,S,1)}\qquad \qquad \qquad \quad \; \;\,$ charm one} \\
\hline
\hline
$(1,-2)$
&$(1,2)$&$K^*$&$-\sqrt2\,g_{\bar36}^9\,h_{99}^9 $\\
\cline{2-3}
&$(1,3)$&$K^*$&$-\sqrt{2}\,g_{66}^9\,h_{99}^9$\\
\cline{2-3}
&$(1,4)$&$D^*$&$-\sqrt{\frac{1}{8}}\,g_{86}^{\bar3}\,h_{\bar39}^{\bar3} - \sqrt{\frac{1}{8}}\,g_{86}^{\bar3}\,h_{9\bar3}^{\bar3}$\\
\cline{2-3}
&$(2,2)$&$\rho$&$-\frac12\,g_{\bar3\bar3}^9\,h_{99}^9 $\\
&&$\omega$&$2\,g_{\bar3\bar3}^1\,h_{99}^9 + \frac12\,g_{\bar3\bar3}^9\,h_{99}^9$\\
&&$\phi$&$-2\,g_{\bar3\bar3}^1\,h_{99}^9 - g_{\bar3\bar3}^9\,h_{99}^9$\\
\cline{2-3}
&$(2,3)$&$\rho$&$-\frac12\,g_{\bar36}^9\,h_{99}^9 $\\
\hline
\hline
\end{tabular}
\caption{Continuation of Tab. \ref{tab:appendix:10}.}
\label{tab:appendix:10b}
\end{center}
\end{table}

\begin{table}
\rescale
\setlength{\tabcolsep}{1.2mm}
\setlength{\arraycolsep}{3.0mm}
\renewcommand{\arraystretch}{1.0}
\begin{center}
\begin{tabular}{|l|l|l|c|}
\hline
\makebox[1cm][l]{$(I,\phantom{x}S)$} & \makebox[1cm][l]{$(a,b)$} & \makebox[1.cm][l]{$V$} & \makebox[9.8cm][l]{$C_{V,ab}^{(I,S,2)}\qquad \qquad \qquad \quad \; \;\,$ charm two} \\
\hline
\hline
$(0,+1)$&$(1,1)$&$\rho$&$-\frac32\,g_{33}^9\,h_{99}^9 $\\
&&$\omega$&$-2\,g_{33}^1\,h_{99}^9 - \frac12\,g_{33}^9\,h_{99}^9$\\
&&$\phi$&$2\,g_{33}^1\,h_{99}^9 + g_{33}^9\,h_{99}^9$\\
\cline{2-3}
&$(1,2)$&$D^*$&$\frac14\,g_{\bar33}^{\bar3}\,h_{\bar39}^{\bar3} + \frac14\,g_{\bar33}^{\bar3}\,h_{9\bar3}^{\bar3}$\\
\cline{2-3}
&$(2,2)$&$\omega$&$-4\,g_{\bar3\bar3}^1\,h_{\bar3\bar3}^1 - 2\,g_{\bar3\bar3}^9\,h_{\bar3\bar3}^1 - 2\,g_{\bar3\bar3}^1\,h_{\bar3\bar3}^9 - g_{\bar3\bar3}^9\,h_{\bar3\bar3}^9$\\
&&$\phi$&$-2\,g_{\bar3\bar3}^1\,h_{\bar3\bar3}^1$\\
&&$J/\Psi$&$-g_{\bar3\bar3}^0\,h_{\bar3\bar3}^0$\\
\hline
\end{tabular}
\caption{Continuation of Tab. \ref{tab:appendix:10b}.}
\label{tab:appendix:11}
\end{center}
\end{table}

\begin{table}
\rescale
\setlength{\tabcolsep}{1.2mm}
\setlength{\arraycolsep}{3.0mm}
\renewcommand{\arraystretch}{1.0}
\begin{center}
\begin{tabular}{|l|l|l|c|}
\hline
\makebox[1cm][l]{$(I,\phantom{x}S)$} & \makebox[1cm][l]{$(a,b)$} & \makebox[1.cm][l]{$V$} & \makebox[9.8cm][l]{$C_{V,ab}^{(I,S,2)}\qquad \qquad \qquad \quad \; \;\,$ charm two} \\
\hline
\hline
$(\frac12,+0)$&$(7,8)$&$\omega$&$-g_{\bar36}^9\,h_{\bar3\bar3}^1-\frac12\,g_{\bar36}^9\,h_{\bar3\bar3}^9 $\\
&&$\phi$&$g_{\bar36}^9\,h_{\bar3\bar3}^1 $\\
\cline{2-3}
&$(7,9)$&$D^*_s$&$\frac14\,g_{\bar33}^{\bar3}\,h_{0\bar3}^{\bar3} + \frac14\,g_{\bar33}^{\bar3}\,h_{\bar30}^{\bar3}$\\
\cline{2-3}
&$(8,8)$&$\omega$&$-4\,g_{66}^1\,h_{\bar3\bar3}^1 - g_{66}^9\,h_{\bar3\bar3}^1 - 2\,g_{66}^1\,h_{\bar3\bar3}^9 - \frac12\,g_{66}^9\,h_{\bar3\bar3}^9$\\
&&$\phi$&$-2\,g_{66}^1\,h_{\bar3\bar3}^1 - g_{66}^9\,h_{\bar3\bar3}^1$\\
&&$J/\Psi$&$-g_{66}^0\,h_{\bar3\bar3}^0$\\
\cline{2-3}
&$(8,9)$&$D^*_s$&$\frac12\,g_{63}^{\bar3}\,h_{0\bar3}^{\bar3} + \frac12\,g_{63}^{\bar3}\,h_{\bar30}^{\bar3}$\\
\hline
\end{tabular}
\caption{Continuation of Tab. \ref{tab:appendix:11}.}
\label{tab:appendix:12}
\end{center}
\end{table}

\begin{table}
\rescale
\setlength{\tabcolsep}{1.2mm}
\setlength{\arraycolsep}{3.0mm}
\renewcommand{\arraystretch}{1.0}
\begin{center}
\begin{tabular}{|l|l|l|c|}
\hline
\makebox[1cm][l]{$(I,\phantom{x}S)$} & \makebox[1cm][l]{$(a,b)$} & \makebox[1.cm][l]{$V$} & \makebox[9.8cm][l]{$C_{V,ab}^{(I,S,2)}\qquad \qquad \qquad \quad \; \;\,$ charm two} \\
\hline
\hline
$(1,-1)$&$(2,2)$&$\rho$&$\frac12\,g_{33}^9\,h_{99}^9 $\\
&&$\omega$&$2\,g_{33}^1\,h_{99}^9 + \frac12\,g_{33}^9\,h_{99}^9$\\
&&$\phi$&$-2\,g_{33}^1\,h_{99}^9 - g_{33}^9\,h_{99}^9$\\
\cline{2-3}
&$(2,3)$&$D^*_s$&$-\sqrt{\frac{1}{32}}\,g_{\bar33}^{\bar3}\,h_{\bar39}^{\bar3} - \sqrt{\frac{1}{32}}\,g_{\bar33}^{\bar3}\,h_{9\bar3}^{\bar3}$\\
\cline{2-3}
&$(2,4)$&$D^*_s$&$-\sqrt{\frac{1}{8}}\,g_{63}^{\bar3}\,h_{\bar39}^{\bar3} - \sqrt{\frac{1}{8}}\,g_{63}^{\bar3}\,h_{9\bar3}^{\bar3}$\\
\cline{2-3}
&$(3,3)$&$\rho$&$-\frac14\,g_{\bar3\bar3}^9\,h_{\bar3\bar3}^9 $\\
&&$\omega$&$-4\,g_{\bar3\bar3}^1\,h_{\bar3\bar3}^1 - g_{\bar3\bar3}^9\,h_{\bar3\bar3}^1 - g_{\bar3\bar3}^1\,h_{\bar3\bar3}^9 - \frac14\,g_{\bar3\bar3}^9\,h_{\bar3\bar3}^9$\\
\hline
\hline
\end{tabular}
\caption{Continuation of Tab. \ref{tab:appendix:12}.}
\label{tab:appendix:13}
\end{center}
\end{table}

\end{document}